\documentclass[
reprint,superscriptaddress,
amsmath,amssymb,aip,floatfix]{revtex4-2}
\usepackage[utf8]{inputenc}
\usepackage{amsmath,amsthm,amssymb,amsfonts,braket,siunitx}
\usepackage{textcomp, gensymb}
\usepackage{placeins}[verbose]
\usepackage{tabularx, longtable, multirow}
\usepackage{graphicx}
\usepackage{dcolumn}
\usepackage{bm,dsfont}
\usepackage[english]{babel}
\usepackage{soul}
\usepackage{comment}
\usepackage[dvipsnames]{xcolor}
\usepackage[normalem]{ulem}
\usepackage[T1]{fontenc}

\newcommand{\sinc}{{\mathrm{sinc}}}

\begin{document}
\title{A High-frequency, Low-power Resonant Radio-frequency Neutron Spin Flipper for High-resolution Spectroscopy}

\altaffiliation{This manuscript has been authored by UT-Battelle, LLC under Contract No. DE-AC05-00OR22725 with the U.S. Department of Energy. The United States Government retains and the publisher, by accepting the article for publication, acknowledges that the United States Government retains a non-exclusive, paid-up, irrevocable, world-wide license to publish or reproduce the published form of this manuscript, or allow others to do so, for United States Government purposes. The Department of Energy will provide public access to these results of federally sponsored research in accordance with the DOE Public Access Plan (http://energy.gov/downloads/doe-public-access-plan).}

\author{Sam McKay}
\altaffiliation{These authors contributed equally to this work
}
\affiliation{Department of Physics, Indiana University, Bloomington, IN 47408, USA}
\affiliation{Center for Exploration of Energy and Matter, Indiana University, Bloomington, IN 47408 USA}
\affiliation{Quantum Science and Engineering Center, Indiana University, Bloomington, IN 47408, USA}

\author{Stephen J. Kuhn}
\altaffiliation{These authors contributed equally to this work
}

\affiliation{Department of Physics, Indiana University, Bloomington, IN 47408, USA}
\affiliation{Neutron Sciences Directorate, Oak Ridge National Laboratory, Oak Ridge, Tennessee 37830, USA}

\author{Jiazhou Shen}
\altaffiliation{Current address: Paul Scherrer Institut, Villigen, Switzerland}
\affiliation{Department of Physics, Indiana University, Bloomington, IN 47408, USA}

\author{Fankang Li}
\affiliation{Neutron Sciences Directorate, Oak Ridge National Laboratory, Oak Ridge, Tennessee 37830, USA}

\author{Jak Doskow}
\affiliation{Center for Exploration of Energy and Matter, Indiana University, Bloomington, IN 47408 USA}

\author{Gerard Visser}
\affiliation{Center for Exploration of Energy and Matter, Indiana University, Bloomington, IN 47408 USA}

\author{Steven R. Parnell}
\altaffiliation{Current address: ISIS, Rutherford Appleton Laboratory, Chilton, Oxfordshire, OX11 0QX, UK}
\affiliation{Department of Physics, Indiana University, Bloomington, IN 47408, USA}
\affiliation{Faculty of Applied Sciences, Delft University of Technology, Mekelweg 15, 2629 JB Delft, The Netherlands.}

\author{Kaleb Burrage}
\affiliation{Neutron Sciences Directorate, Oak Ridge National Laboratory, Oak Ridge, Tennessee 37830, USA}

\author{Fumiaki Funama}
\affiliation{Neutron Sciences Directorate, Oak Ridge National Laboratory, Oak Ridge, Tennessee 37830, USA}

\author{Roger Pynn}

\affiliation{Department of Physics, Indiana University, Bloomington, IN 47408, USA}
\affiliation{Center for Exploration of Energy and Matter, Indiana University, Bloomington, IN 47408 USA}
\affiliation{Quantum Science and Engineering Center, Indiana University, Bloomington, IN 47408, USA}
\affiliation{Neutron Sciences Directorate, Oak Ridge National Laboratory, Oak Ridge, Tennessee 37830, USA}

\date{\today}
\begin{abstract}
We present a resonant-mode, transverse-field, radio-frequency (rf) neutron spin flipper design that uses high-temperature superconducting films to ensure sharp transitions between uniform magnetic field regions.
Resonant mode allows for low power, high frequency operation but requires strict homogeneity of the magnetic fields inside the device. This design was found to efficiently flip neutrons at 96.6$\pm 0.6\%$ at an effective frequency of 4 MHz with a beam size of $2.5~\times~2.5$~cm and a wavelength of 0.4 nm.
The high frequency and efficiency enable this device to perform high-resolution neutron spectroscopy with comparable performance to currently implemented rf flipper designs. The limitation of the maximum frequency was found due to the field homogeneity of the device. We numerically analyze the maximum possible efficiency of this design using a Bloch solver simulation with magnetic fields generated from finite-element simulations. We also discuss future improvements of the efficiency and frequency to the design based on the experimental and simulation results.
\end{abstract}

\maketitle

\section{Introduction}

Neutron Resonance Spin Echo (NRSE) is a high energy-resolution experimental technique that is commonly used to probe the dynamics of both hard and soft condensed matter systems. \cite{Gardner_2020,Keller_2022}
While the closely-related technique of Neutron Spin Echo (NSE) uses strong static magnetic fields to generate neutron spin precession, NRSE instead uses radio-frequency (rf) neutron spin flippers separated by a zero-field region. In general, rf flippers manipulate the neutron polarization by using an rf field tuned to a resonant frequency determined by the strength of a (typically stronger) static magnetic field.
The flipping efficiency of the rf flipper largely determines the performance of the NRSE instrument, an important factor in deciding if it is an improvement over a similar NSE instrument. \cite{Golub1987} In order to make NRSE competitive with NSE, an rf flipper must: run at a high frequency ($\sim$5 MHz), flip the neutron spin with a high efficiency ($\gtrsim 95\%$), and be easily tuned and reliably run for multiple days during an experiment.

Broadly speaking, there are two methods of spin manipulation that rf flippers use, each involving both a static field and an oscillating rf field perpendicular to the static field.
The first method is the \textit{adiabatic mode} (also called gradient mode), which applies an additional static-field gradient across the flipper along with a large-amplitude rf field. In this method, the neutron spin direction can be considered to follow the gradient field where it is strong and the rf field where the Larmor frequency determined by the static field is close to the rf frequency.\cite{Grigoriev2001,Li2020} Thus, the neutron spin adiabatically follows the field direction throughout its path through the flipper.
The second method is the \textit{resonant mode} (also called non-adiabatic mode), where the rf field magnitude is set much lower and the neutron has a high likelihood of flipping (via single-photon exchange) which changes the total energy of the neutron.\cite{SUMMHAMMER1993}
While the effect on the neutron spin and energy is essentially the same for both modes, the technical limitations of rf flippers based on these methods are quite different. In the adiabatic mode, the rf field is much stronger and an extra gradient field coil must be installed, leading to a much higher power output of the flipper. In the resonant mode, the power is lower but the efficiency is much more sensitive to small deviations in the fields, so a higher field homogeneity of both the static and the rf fields is required.
These factors have limited the maximum frequency of adiabatic mode rf flippers to 2 MHz \cite{Li2020} and the maximum efficiency of resonant mode rf flippers at high frequency as well.\cite{Kitaguchi2009, Franz2019}

In this paper, we present a prototype resonant rf flipper that uses high-temperature superconducting (HTS) films as Meissner screens to enhance the field homogeneity by reducing the effects of fringe and stray magnetic field, so low power output is essential in order to be able to cool the device when run at high frequencies. The device consists of two rf flippers in a single vacuum chamber that can work additively in the so-called \textit{bootstrap mode} to double the effective frequency. \cite{KellerTbootstrap} Each flipper consists of a pair of static-field coils, an rf coil that produces a longitudinal field along the beam, and HTS films that sharply define the magnetic field regions. In bootstrap mode, the device produces an effective frequency of 4~MHz with a flipping efficiency of 96.6\% and a beam size of about $2.5~\times~2.5$~cm at a wavelength of 0.4 nm.
To explore the efficiency of our design, we used a numerical Bloch solver that simulates the evolution of the neutron spin as it passes through the rf flipper.
By analyzing the effect of the static and rf field inhomogeneities inside of our prototype device, we propose future improvements to our resonant rf flipper design.

\section{The Resonance Condition} \label{sec:theory}

\begin{figure}[b]
\includegraphics[width=.95\linewidth]{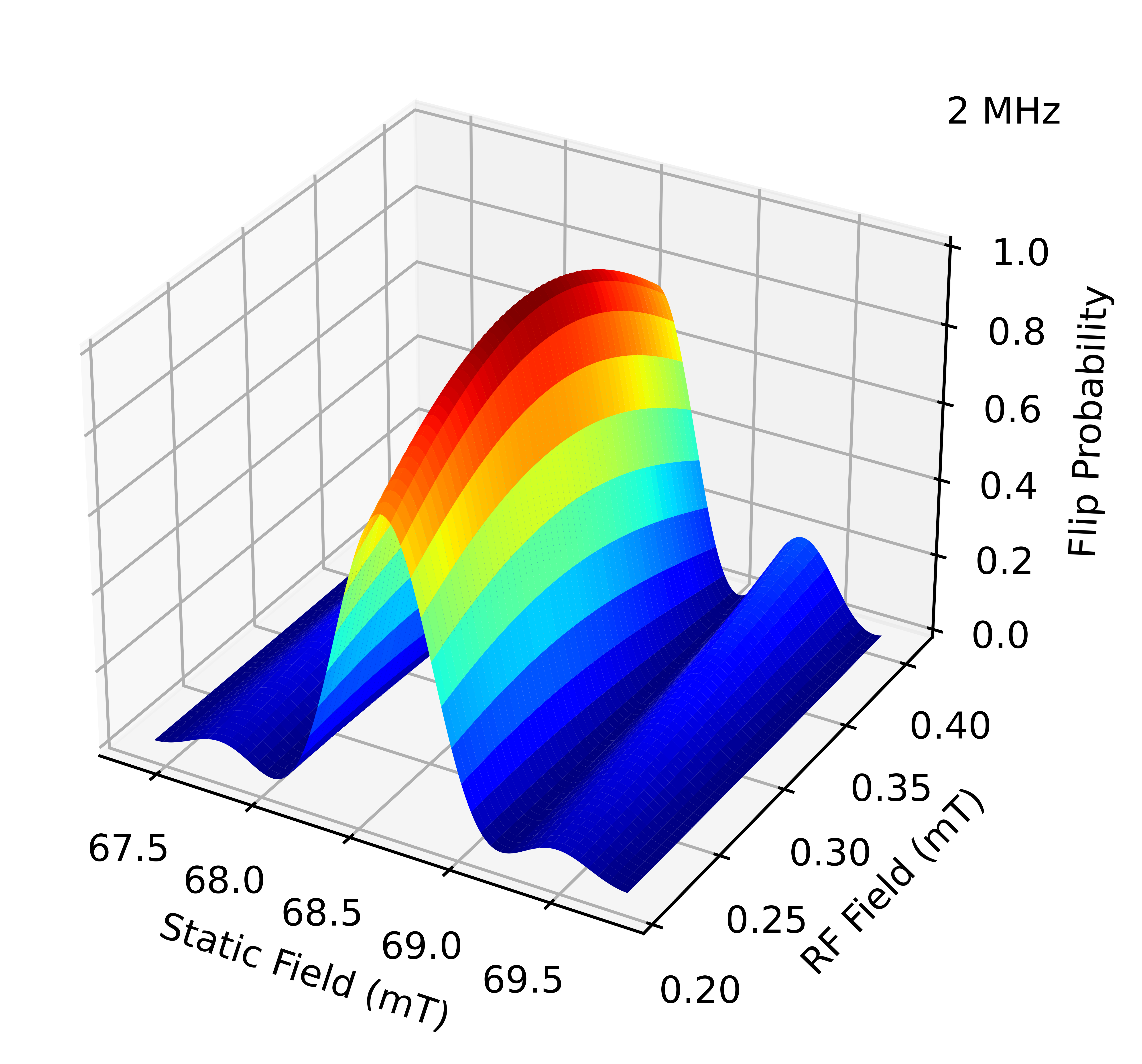}
\caption{\label{figResonanceCurve} An example flip probability given in Eqn. \eqref{eq:spin flip prob} as a function of static and rf field strength with an rf frequency 2 MHz, flipper length of 8 cm, and neutron wavelength of 0.55 nm. The probability reaches unity when the resonance conditions shown in Eqn. \eqref{eq:res condition} are satisfied. Notice that the flipping probability is much sharper with respect to the static field than the rf field.}
\end{figure}

In this section, we review the standard mathematical theory that has been shown to well-describe the action of rf flippers; for a more extensive review, see Cook. \cite{Cook_2014}
The Hamiltonian that describes our rf flipper is given in the lab frame by
\begin{equation}
\begin{aligned} \label{eq:lab Ham}
    \mathcal{H}_{\mathrm{lab}}(t) = \frac{|\gamma|\hbar}{2} \Big[& \sigma_x B_{\mathrm{rf}}(t,\bm{r}(t)) \cos(\omega t + \phi_0)   + \\
    &\sigma_z B_0(t,\bm{r}(t)) \Big],
\end{aligned}
\end{equation}
where $\gamma \approx \SI{-1.832e8}{\radian\per\tesla\per\second}$ is the gyromagnetic ratio of the neutron, $\hbar$ the reduced Planck's constant, $\sigma_k$ for $k \in \{ x,y,z\}$ the usual Pauli matrices, $\bm{r}(t)$ the classical path of the neutron, $\omega = 2 \pi f$ the angular rf frequency, and $\phi_0$ the phase of the rf field. Here, we explicitly split the magnetic field into a static field component $B_0$ and and rf amplitude envelope $B_{\mathrm{rf}}$, both of which may be time-dependent.
To better understand the dynamics, we consider the Hamiltonian in the coordinate system that rotates about the direction of the static magnetic field at a frequency equal to the rf frequency. Mathematically, this transformation is accomplished by the replacement $\ket{\psi}_{\mathrm{rot}} = e^{-i (\omega t +\phi_0) \sigma_z /2 } \ket{\psi}_{\mathrm{lab}}$.
Performing the substitution, we obtain
\begin{equation}
\begin{aligned} \label{eq:rot Ham}
    \mathcal{H}_{\mathrm{rot}}(t) =& \sigma_x \frac{|\gamma| \hbar}{4} B_{\mathrm{rf}}(t) \left[ 1 + \cos(2 \omega t + \phi_0)\right] + \\
    &\sigma_y \frac{|\gamma| \hbar}{4} B_{\mathrm{rf}}(t) \sin(2 \omega t + \phi_0) - \\
    &\sigma_z \frac{\hbar}{2} \left[\gamma B_0(t) - \omega \right],
\end{aligned}
\end{equation}
where for clarity we have suppressed the dependence of the magnetic field on the classical path of the neutron through the instrument.

We first consider the case where the magnetic field coefficients are \textit{time-independent} after averaging over many cycles of the rf signal. The time-averaged Hamiltonian in the rotating frame is given by
\begin{equation}
    \label{eq:tav rot Ham}
    \overline{\mathcal{H}}_{\mathrm{rot}} = \frac{\hbar}{2}\left[\sigma_x |\gamma| (\overline{B}_{\mathrm{rf}}/2)
    - \sigma_z \left(\gamma \overline{B}_0 - \omega \right)\right]
\end{equation}
where the bar represents time-averaging.
The solution to the Schrodinger equation with  Eqn. \eqref{eq:tav rot Ham} leads to the well-known Rabi equation \cite{Rabi1937rf} for the spin-flip probability:
\begin{subequations} \label{eq:spin flip prob}
\begin{align}
    P^{\mathrm{(id)}}_f(t) &= \frac{(\gamma \overline{B}_{\mathrm{rf}}/2)^2}{\omega_r^2} \sin^2 \left( \frac{\omega_r t}{2} \right), \\
     \omega_r &= \sqrt{(\gamma \overline{B}_0 - \omega)^2 + (\gamma \overline{B}_{\mathrm{rf}}/2)^2},
\end{align}
\end{subequations}
where $\omega_r$ is the Rabi frequency for our Hamiltonian in Eqn. \eqref{eq:tav rot Ham}. 
From Eqns. \eqref{eq:tav rot Ham} and \eqref{eq:spin flip prob}, we see that there are two conditions for an efficient spin flip: the rf frequency must equal the \textit{Larmor frequency} $\gamma \overline{B}_0$ for the static magnetic field, and the rf amplitude must be tuned to the neutron travel time through the flipper.
These two conditions for resonance are given by the equations
\begin{subequations} \label{eq:res condition}
\begin{align} 
    \overline{B}^{\mathrm{(res)}}_0 &= \frac{\omega}{\gamma}, \\
    \overline{B}^{\mathrm{(res)}}_{\mathrm{rf}} &= \frac{2 \pi}{\gamma T} = \frac{(2 \pi)^2 \hbar}{\gamma m L \lambda},
\end{align}
\end{subequations}
where $T$ is the total transit time of the neutron through the rf flipper, $L$ the length of the flipper, and $\lambda$ the wavelength of the neutron.

Although the resonance condition has been derived in the case of an ideal magnetic field profile, similar resonance conditions apply to nonuniform field profiles (see Sec. \ref{sec:QBS}). However, field inhomogeneities will change the shape of the resonance tuning curve: if the field profiles are too irregular, then the resonance curve will be too sharp, thus impacting the practical flipping efficiency of the rf flipper. One must also take into account thermal fluctuations, as they may change the resistance in the coils and therefore the magnetic field profile. In general, the flipping efficiency is dependent on the homogeneity of the static field and the stability of the rf field. \cite{Williams1998,Martin_2014}
As detailed in the next section, our approach to the field uniformity problem involves using HTS coils and HTS films in order to generate a stable, well-contained rf field and strong, uniform static field inside the rf flipper.

\section{Design and Simulation} \label{sec:design}

\begin{figure}[ht]
\includegraphics[width=.9\linewidth]{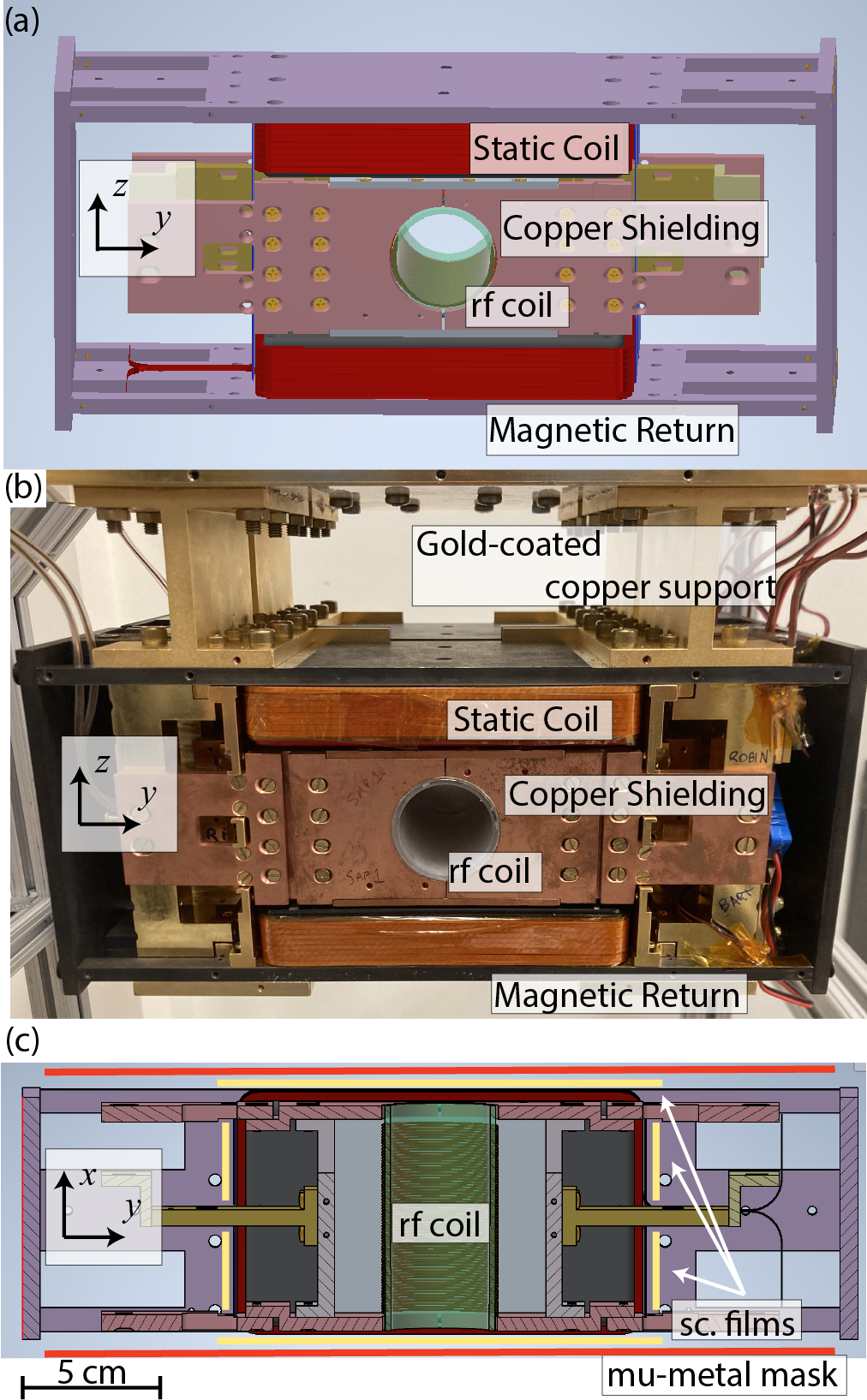}
\caption{\label{figRFinstrument} (a)~Drawing of the front view of the magnetically-relevant components of the rf flipper. The neutron beam travels through the longitudinal rf coil at the center of the device (in the $x$-direction). A box of copper shielding surrounds the coil to screen the rf field. The upper and lower static coils, made of high-temperature superconducting (HTS) wire, produce a vertical field through the rf coil region. An iron magnetic return loop surrounds the device to homogenize the static field. The HTS films and mu-metal shielding are not shown. (b)~The front view with the components from (a) as well as the structural components of a sapphire tube supporting the rf coil and copper pieces connecting the components together. Two rf flippers are pictured, one behind the other; this entire assembly is placed inside of a single vacuum chamber, allowing the device to be operated in bootstrap mode. (c)~Drawing of the equatorial slice through a flipper. The 6 HTS films that provide a sharp field transition and homogenize the field are shown in yellow. A mu-metal mask to screen stray field is shown in red in front of and behind the rf flipper. }
\end{figure}

The design process of the rf flipper is naturally divided into two sections, namely the design of the components that generate the rf field and static field. These fields were independently simulated via the finite-element method with the Siemens MagNet $\copyright$ software, which includes the material properties in its solutions. The components of the device are shown in Fig. \ref{figRFinstrument}.

\subsection{Static Field}

\begin{figure}[ht]
\includegraphics[width=.9\linewidth]{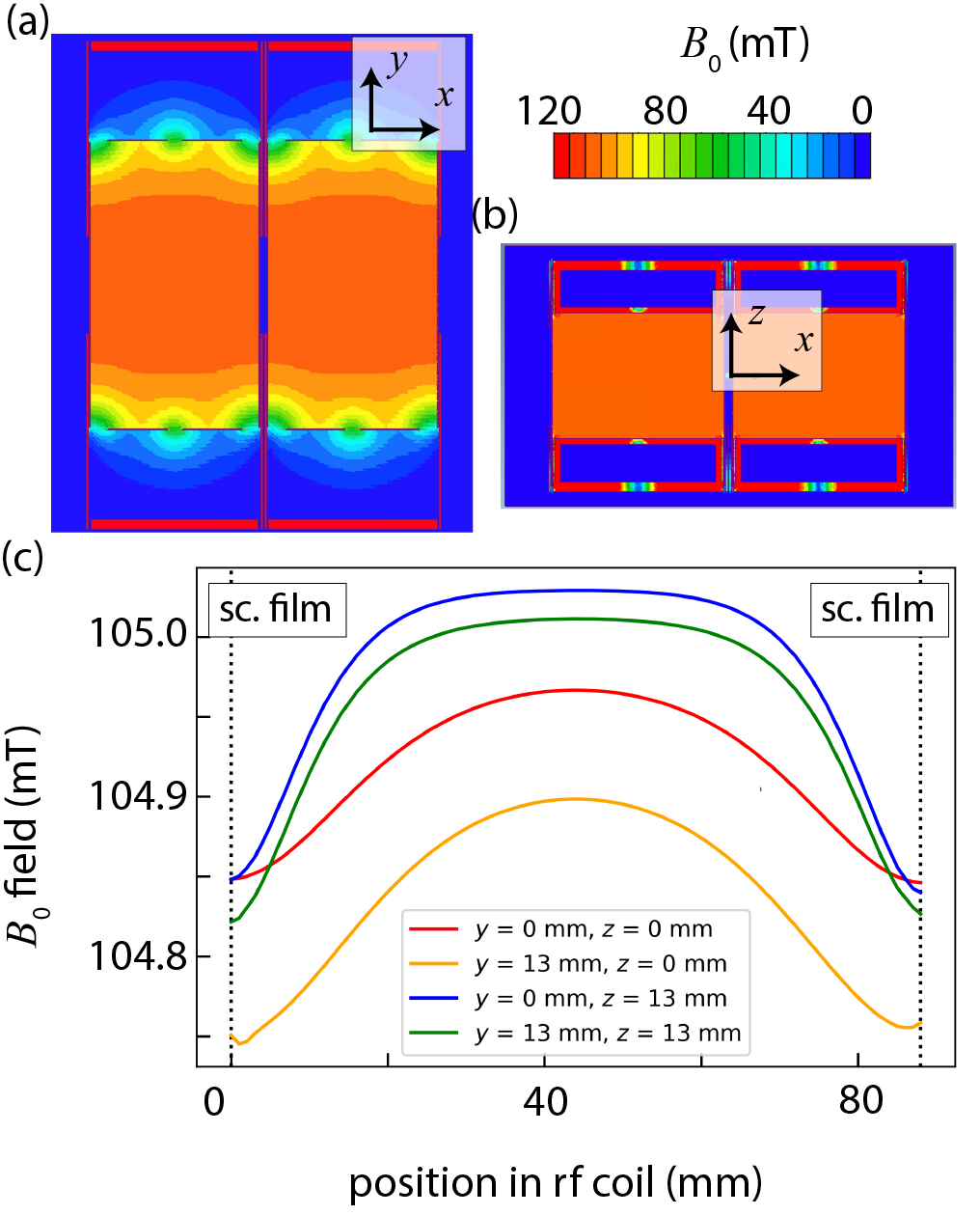}
\caption{\label{figStaticFields} MagNet finite-element simulations of the static field in the (a) horizontal ($x$-$y$) plane and the (b) vertical ($x$-$z$) plane. The neutron beam goes from left to right through the center of the device. The field is well contained by the high-temperature superconducting (HTS) films at the front back and sides of the device. The magnetic flux returns and mu-metal shielding at the top, bottom, and edges have the highest field. The $x$ and $y$ components of the static field were found to be negligibly small, so $\bm{B}_0 \approx B_0 \hat{z}$.
(c) The simulated field for neutrons travelling through the device at various distances from the beam center, with the HTS films at $x = 0.87 \ \mathrm{mm}$.}
\end{figure}

The vertical static field is produced by a pair of coils wound with HTS tape (Superpower 2G HTS wire) around low carbon steel (alloy 1018) hollow pole pieces, as shown in Fig. \ref{figRFinstrument}. The static field was designed to have the maximum homogeneity in the beam region. The pole pieces were 14 cm wide, much wider than the expected beam size of $2.5 \times 2.5$ cm, in order to increase field uniformity. 
A magnetic flux return, also made of low carbon steel, completes the magnetic circuit away from the neutron beam. The homogeneity of the static field was further increased by adding HTS films (Ceraco ceramic coating GmbH) on the front, back and sides of the coils.
Here `front' refers to the film the neutron will encounter before entering the flipper. The 350 nm thick YBa$_2$Cu$_3$O$_7$ (YBCO) films are deposited on a 0.5 mm sapphire plate and coated with 100 nm of gold to prevent oxidation. At the operating temperature of 25 K and at fields below $H_{c1}$, \cite{ybcoHc1} the films are in the Meissner superconducting state, which constrains the field and also enhances the homogeneity. \cite{Wang2014}
The side HTS films have gaps for the rf shield support beams in the front, middle, and back, as shown in Fig. \ref{figRFinstrument}(c). MagNet simulations found that these gaps at the edge have little effect on the field homogeneity in the beam because they are far from the beam center. High permeability mu-metal plates were installed at the front and back of each flipper to prevent external magnetic fields from penetrating into the device and also to prevent the internal static field from affecting the guide fields outside the device. MagNet simulations found that this stray field was about $0.01\%$ of the field at the center of the device, even with a cut-out in the center to allow the beam to pass through. The maximum height and width of the static field region of the device was determined by the maximum size of available YBCO films (currently $148 \times 112$ mm). 

The $B_0$ coils consist of 64 turns of HTS wire with a maximum rated current of 50 amps. MagNet simulations find that this produces a field of up to approximately 110~mT in the $2.5 \times 2.5$ cm beam area, as shown in Fig.~\ref{figStaticFields}(a,b). MagNet simulations have been experimentally verified in previous static field devices. \cite{Wang2014} The simulated field is uniform across the beam to within 0.25 mT, as shown in Fig. \ref{figStaticFields}(c). Notably, the variation is greatest closest to the front and back HTS films, consistent with the results from similar simulations. \cite{Li2020rfsim} 

\begin{figure}
\includegraphics[width=.95\linewidth]{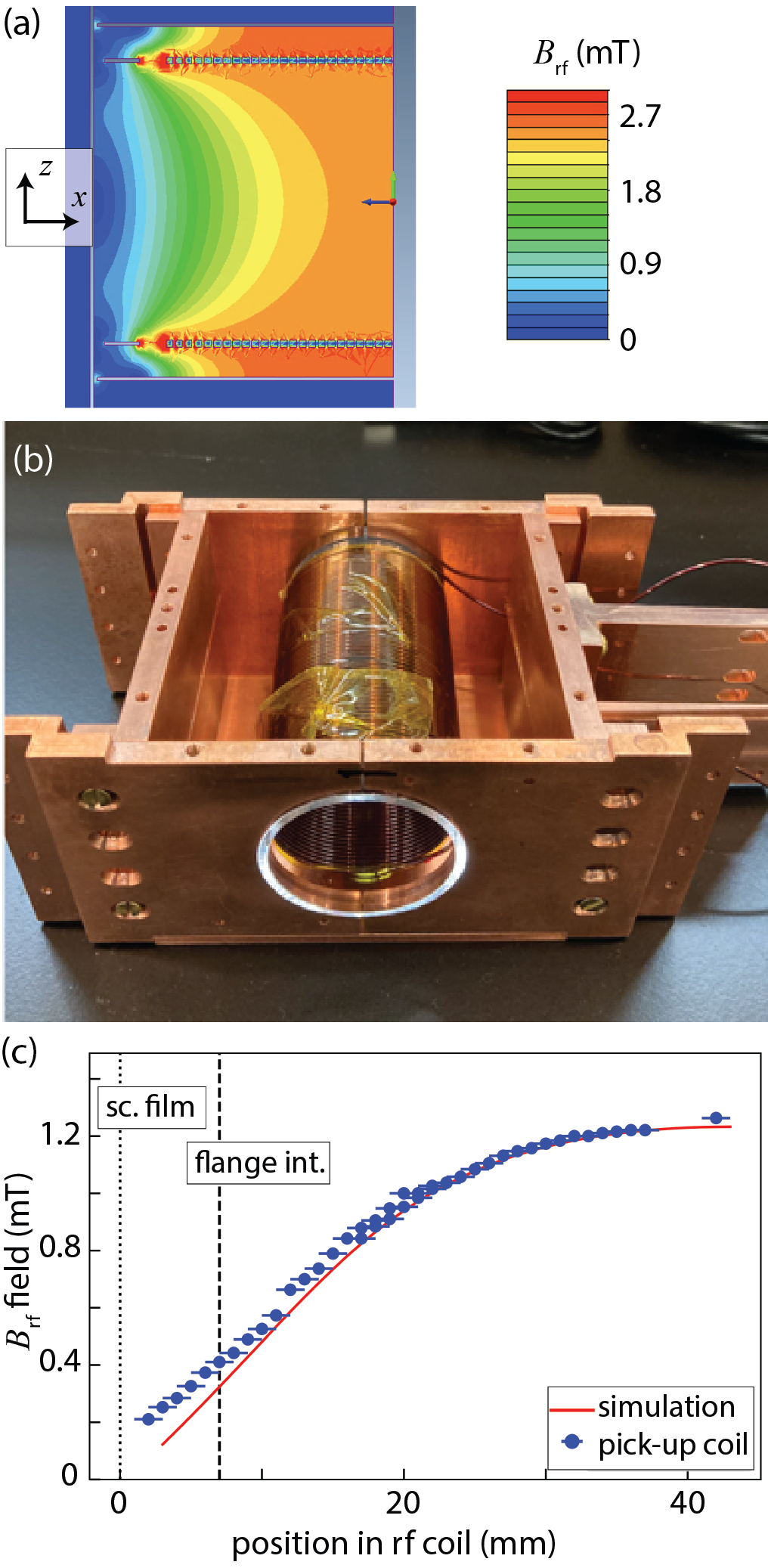}
\caption{\label{figRFdesign} (a)~Simulations of the field in the vertical ($x$-$z$) plane from the rf coil and copper shielding (only half the coil length is simulated). The beam travels from the left to right, passing through the HTS film and through the center of the solenoid. (b)~Aluminum wire wrapped around sapphire creates the rf field. The copper flange holds the sapphire tube and provides rf shielding. (c)~Comparison of the simulated field from (a) to the field measured with a pick-up coil along the axis of the sapphire cylinder (the $x$-axis). The HTS film is at $x = 0$ mm and the interior of the copper shielding flange is at $x = 7$ mm. The HTS film was included in the simulations but not installed for the pick-up coil measurement.}
\end{figure}

\begin{figure}
\includegraphics[width=.95\linewidth]{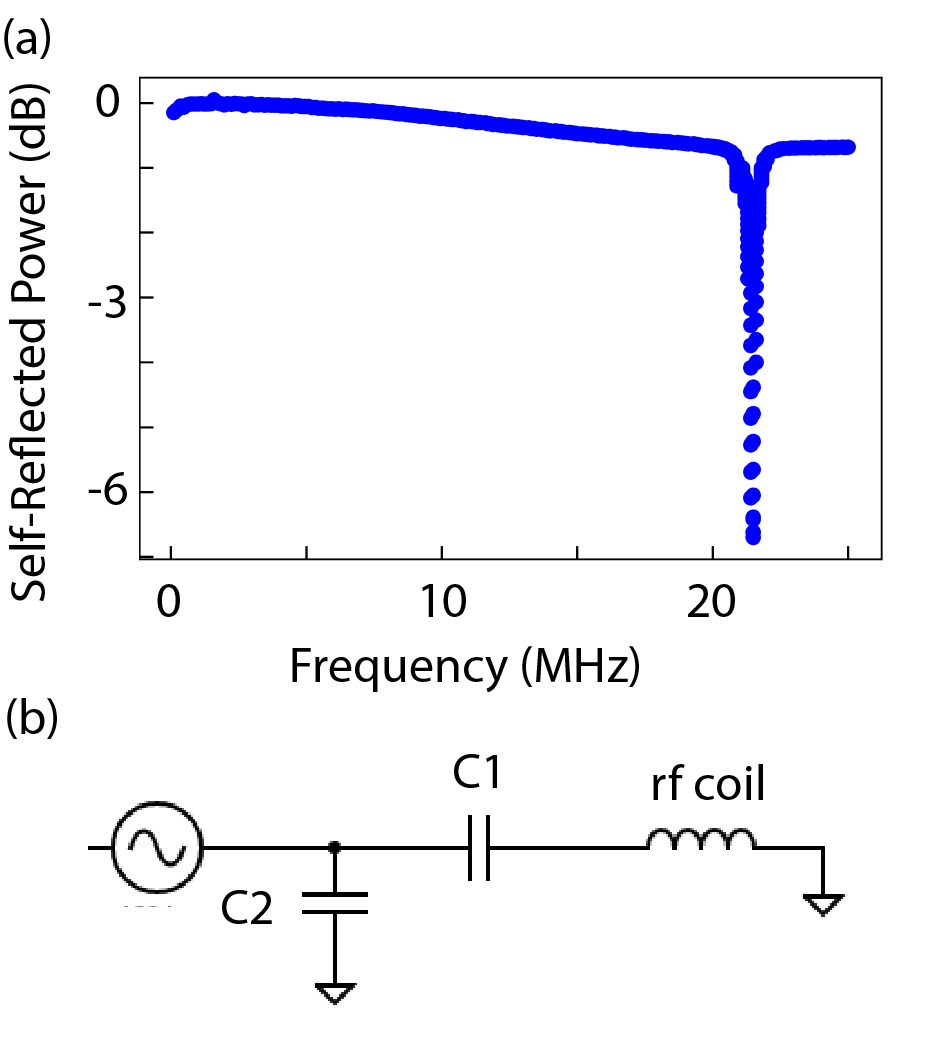}
\caption{\label{figRFcircuit} (a)~The rf circuit self-reflected power measurement (s11) of the unmatched rf coil vs. frequency at 22 K. (b)~The schematic of the matching circuit used to create a resonance at the desired frequency with an impedance of $\sim 50 \ \Omega$ to match the rf amplifier output. Capacitor values for some rf frequencies are given in Tab.~\ref{tab:Capacitors}.}
\end{figure}

\subsection{Radio-frequency Field}

The rf field is produced by a cylindrical solenoid, with the beam passing through the center. The solenoid is wound with 57 turns of 18~gauge aluminum wire around a 4 cm outer diameter, 8.1~cm long sapphire tube. Sapphire was chosen for its good thermal conductivity and superior strength, and aluminum wire was chosen for its ease of winding. The wire is held to the sapphire tube with ultraviolet glue (Norland Optic 61) and Kapton tape.
A copper box to shield the rf frame was designed to encase the rf coil. Copper flanges on the front and back support the sapphire tube, as shown in Fig. \ref{figRFdesign}. The dimensions of the rf shielding were chosen to be as large as possible without touching the pole pieces or YBCO film in order to prevent heating or damaging the delicate YBCO; there is a $\sim 1$ cm gap between the coil and shield box at its closest point.

The copper flanges on the front and back have a thickness of $\sim 0.6$ cm, which was chosen to ensure good thermal contact with the sapphire tube as well as to reduce the rf field strength at the front and back of the device in the beam region. 
The flanges reduce the field strength to $\sim 1/6$ of the maximum field strength at the gold-coated film position, according to both the MagNet simulation and pick-up coil test shown in Fig. \ref{figRFdesign}(c). By reducing the rf field strength close to the films, the gold-coated YBCO films function more effectively as rf field shielding. The rf shielding effectiveness of the YBCO films was checked by powering the first rf flipper (rf1) and placing a pick-up coil in the center of the second rf flipper (rf2). The field measured by the pick-up coil was found to be at most $1.2\%$ of the powered field. When the YBCO films were replaced by aluminum sheets, the pick-up coil response was $0.2\%$ or less, suggesting that the gold-coated YBCO films are not perfect rf shields in the operating frequency range.
The simulated rf field uniformity is shown in Fig. \ref{figRFdesign}(a). The ratio of the rf field integral along the center to the field integral at radius of 13 mm is 90\%. The simulated field was experimentally confirmed by translating a pick-up coil through the rf coil optic axis at room temperature with the front and back films removed, as shown in Fig. \ref{figRFdesign}(c). The simulation includes the YBCO film, while the pick-up coil measurement does not, which explains the discrepancy at the edge of the coil.

The rf circuit for each independently-driven rf coil consisted of a function generator, rf amplifier, matching circuit, and rf coil. A self-reflected power measurement (s11 measurement \cite{Huang2008}) was conducted using a network analyzer, as shown in Fig. \ref{figRFcircuit}(a). If the reflected power is close to 0~dB, then almost all the power is reflected back into the transmitter, meaning that the circuit is not in resonance by some unintentional coupling of the components of the rf coil, shielding and leads. The unmatched rf coil at 22~K has little self-resonance up to 20~MHz.
The rf coil produces a field of $\sim$0.6 mT/ amp at 2 MHz. Different matching circuits each consisting of two capacitors (labeled C1 and C2) were used for each rf frequency. As shown in Fig. \ref{figRFcircuit}(b), C1 is used to put the capacitor/rf coil circuit on resonance and C2 is used to impedance-match the circuit to the 50~$\Omega$ output of the rf amplifier; see Tab. \ref{tab:Capacitors} for specific matching circuit capacitances. Because the power output was quite low, the matching circuit impedance only needed to be in the range of 10-90~$\Omega$. In order to reduce the self-resonance in the leads, the matching circuits were found to perform best when attached directly to the outside of the vacuum chamber feedthroughs.

Good thermal contact is essential between the rf coil and the Sumitomo CH-110 dry closed-cycle refrigerator with water-cooled compressor (CCR) used to cool the assembly. The thermal contact was improved by using sapphire for the rf coil structure, compressing indium into the junction of the sapphire tube and the copper rf frame, and using copper 101 (oxygen free) for the connection pieces between the CCR and the rf coil. With these measures, the device was held at a stable temperature, $\sim$5 K above the base temperature of 21 K, even with the rf coils powered to 10 Watts (1 rf coil powered at 10 Watts and both rf coils powered at 5 Watts each were found to have approximately the same heating). 
In operation, a power output of $\sim$1~Watt per coil is sufficient, so the flipper could be run using a water-cooled compressor (Sumitomo F-70L base temperature 21 K) or air-cooled compressor (Sumitomo HC-4A base temperature 30 K). A vacuum of $\sim 10^{-7}$~Torr and a thin pure aluminum heat shield around the flux returns kept the devices from being warmed by the environment. 

\begin{table}[t]
\centering
\newcolumntype{R}{>{\centering\arraybackslash}X}
\begin{tabularx}{.95\linewidth}{R|R|R}
       Frequency (MHz) &
       C1 (nF) &
       C2 (nF) 
       \\
\hline
0.56 & 1.8 & 39 \\
2.02 & 0.1 & 5.6 \\
2.53 & 0.05 & 3.9 \\
2.95 & 0.03 & 1.5 \\
\end{tabularx}
\renewcommand{\arraystretch}{1}
\caption{\label{tab:Capacitors} 
Capacitor values for the matching circuits at different frequencies shown in Fig. \ref{figRFcircuit}.}
\end{table}

\section{Quantum Bloch Solver Simulations} \label{sec:QBS}

\begin{figure}[b]
\includegraphics[width=.95\linewidth]{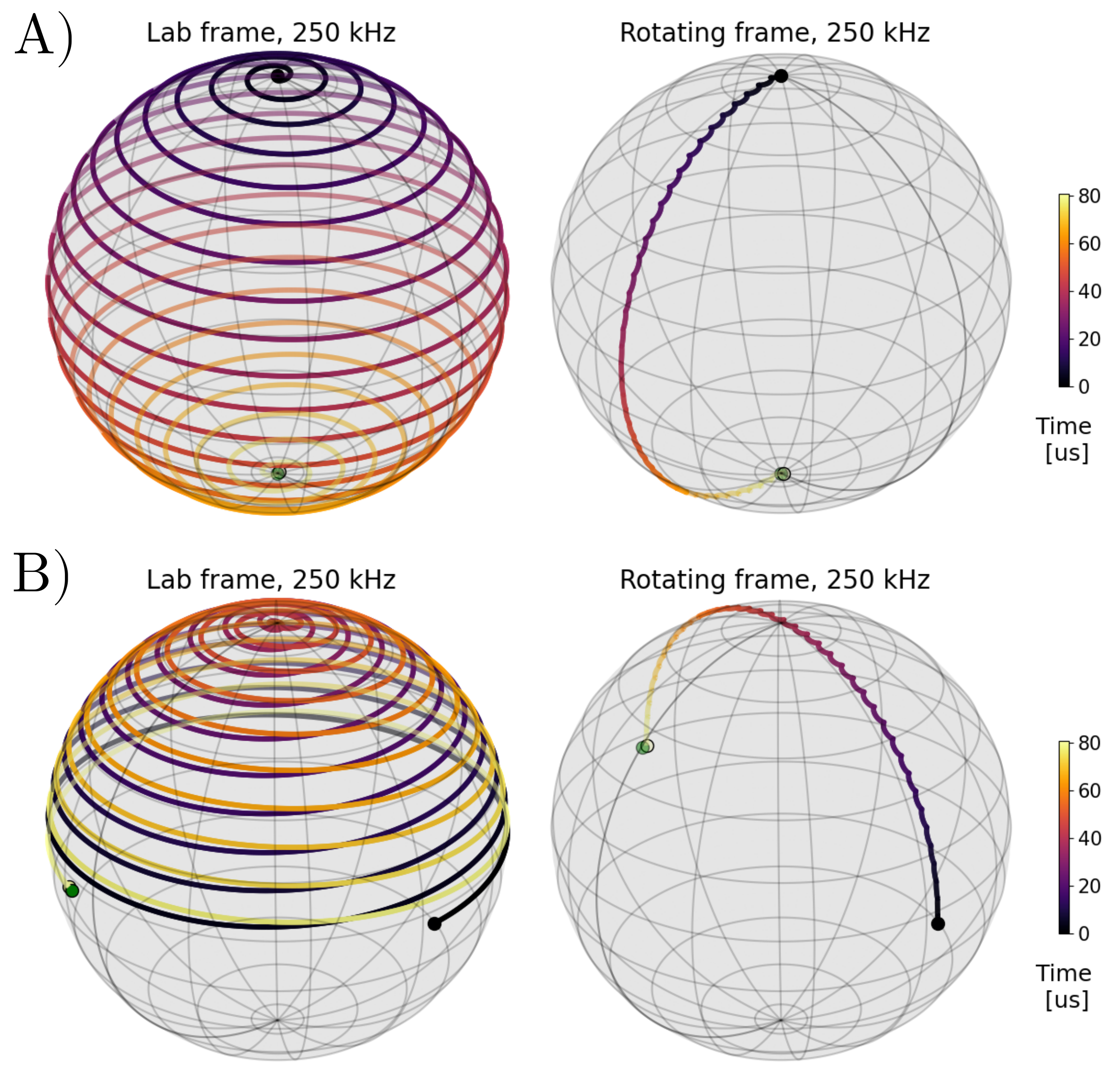}
\caption{\label{fig:labrotframe} A) Example flipping mode and B) precession mode calculated neutron polarization evolution through the rf flipper at 250 kHz for a single trajectory using the static and rf magnetic fields from the MagNet finite-element simulations. The green dot on each of the Bloch spheres is the goal final polarization for that mode. The low frequency was chosen to clearly display the trajectory of the polarization on the Bloch sphere.}
\end{figure}

In order to understand the qualitative effects of a non-uniform static field and rf oscillation envelope on the efficiency of the rf flipper, we performed Bloch-solver simulations (i.e., numerically solving the Bloch equations\cite{Bloch_1946}) of various field models in order to understand design limitations and explain the experimental results reported later in this work. We also analyzed the flipping efficiency using the magnetic fields output by MagNet simulations as discussed in Sec. \ref{sec:design}. For more discussion on the method used in these simulations, see Appx. \ref{appx:qbs}.
Because of the boundary conditions due to the HTS films as discussed in Sec. \ref{sec:design}, we assume throughout this section that the rf envelope vanishes at the flipper boundary.

The 3D real-valued neutron polarization vector $\bm{P}_{\psi}$ for the spinor $\ket{\psi}$ is defined in the usual way: $\bm{P}_{\psi}~=~(\langle \sigma_x \rangle,\langle \sigma_y \rangle,\langle \sigma_z \rangle)^T$ where the angle brackets represent the expectation over the state $\ket{\psi}$. We only consider pure states in this work, i.e. states that exist on the surface of the Bloch sphere, so $|\bm{P}_{\psi}| = 1$.
We define the generalized efficiency $\varepsilon$ of the rf flipper as 
\begin{equation} \label{eq:general eff}
    \varepsilon = \frac{1 + \bm{P}_f \cdot \bm{P}_g}{2}, 
\end{equation}
where $\bm{P}_f$ and $\bm{P}_g$ are the final and ``goal'' polarization states, respectively.  
The final polarization goal is different for \textit{precession mode} and \textit{flipping mode}: in flipping mode, the neutron enters the device parallel or antiparallel to the static field as shown in Fig. \ref{fig:labrotframe}(a) and the desired final state is the antipodal point on the Bloch sphere given by $\bm{P}_g = -\bm{P}_i$ where $\bm{P}_{i}$ is the initial polarization; in precession mode shown in Fig. \ref{fig:labrotframe}(b), the initial polarization state enters the flipper in the plane perpendicular to the static field, and the desired final state corresponds to a return to the equator of the Bloch sphere  with a final neutron azimuthal phase of
\begin{equation} \label{eq:rf flipper phase}
    \phi_f = \omega(T + 2 t_0) - \phi_i,
\end{equation}
where $t_0$ is the entrance time of the neutron and $\phi_i$ the initial neutron phase. In terms of the polarization, this phase change corresponds to a goal final polarization of
\begin{equation}
    \bm{P}_g = \bm{R}_{\hat{n}}(\phi_f) \bm{P}_i,
\end{equation}
where $\bm{R}_{\hat{n}}(\cdot)$ is the appropriate rotation matrix about the axis $\hat{n}$ defined by the static field.

\begin{figure}[tb]
\includegraphics[width=.9\linewidth]{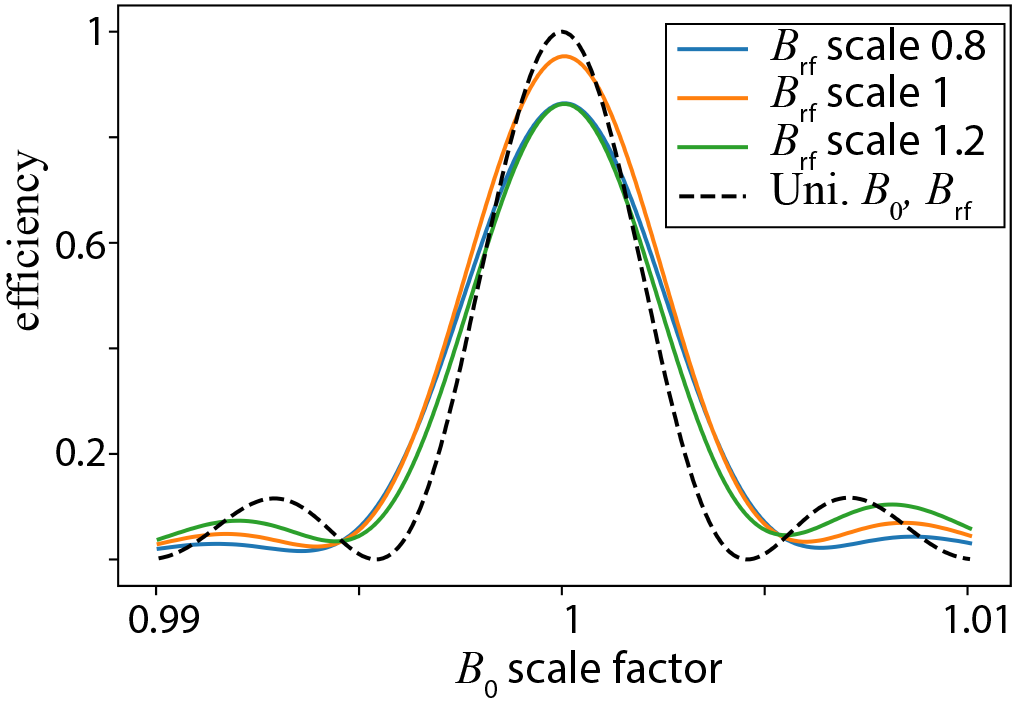}
\caption{\label{fig:ShoulderExample} Example of the simulated efficiency of the flipper of the whole beam as a function of the static field strength and rf field strength at 2 MHz. The simulated fields are multiplied by a scale factor, equivalent to performing a current scan in the real device. The dashed curve represents a device with strictly uniform static field and rf field amplitude throughout the device which corresponds exactly to the Rabi equation for spin-flip probability given in Eqn.~\eqref{eq:spin flip prob}. }
\end{figure}

\begin{figure}[ht]
\includegraphics[width=.95\linewidth]{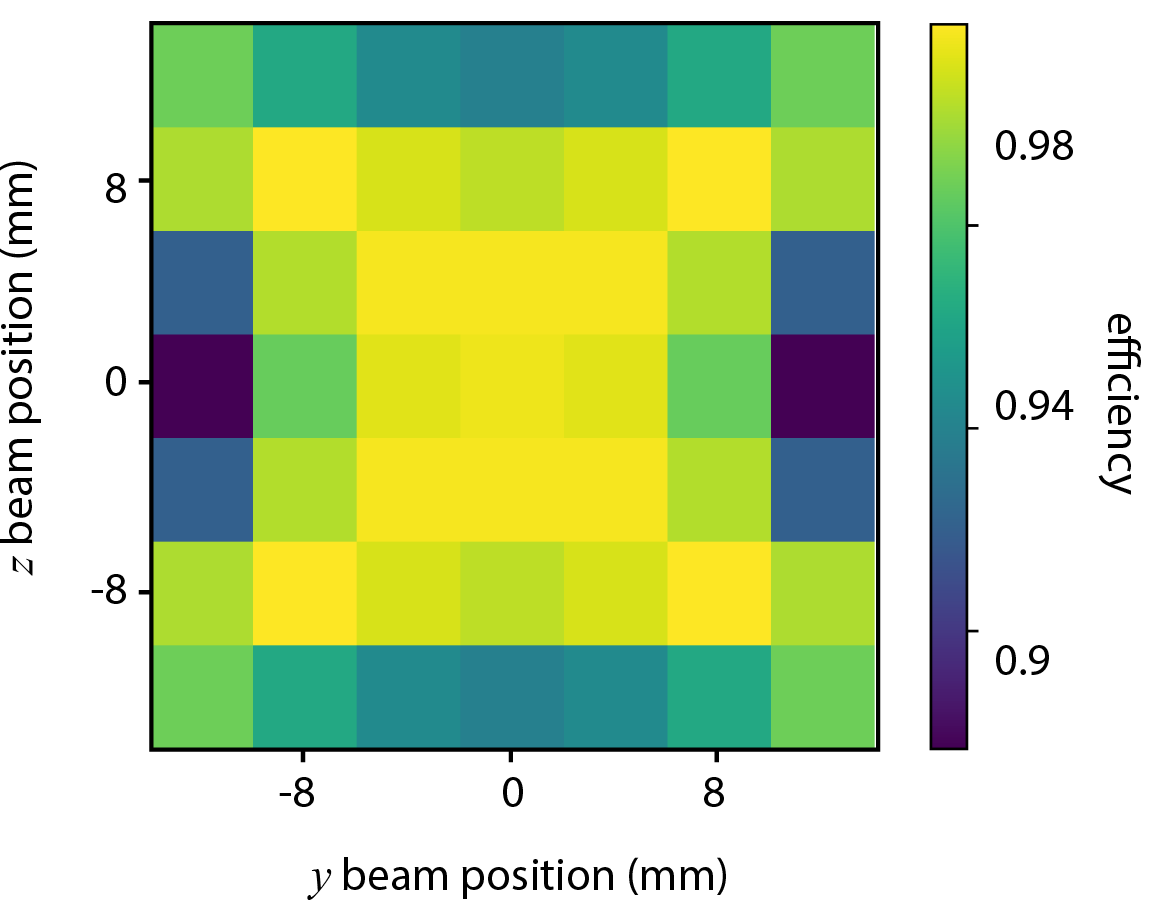}
\caption{\label{fig:simbeamefficiency} Simulated efficiency of the rf flipper at 2 MHz in flipping mode with the simulated static and rf fields for different trajectories parallel to the optic axis. The efficiency is defined by Eqn.~\eqref{eq:general eff}.}
\end{figure}

We now turn to simulating the rf flipper magnetic field models determined by MagNet. Just one coil of the bootstrap pair was simulated, so a simulation at 2 MHz reflects the efficiency of a coil in the effective 4 MHz bootstrapped set-up. A beam size of 24 $\times$ 24 mm was simulated in 4 mm pixels of $B_{\mathrm{rf}}(x,y,z)$ and $B_{\mathrm{0}}(z)$; only the positive $y$ and $z$ values were simulated as the device is symmetric around the $y$ and $z$ axes, and $B_{\mathrm{0}}(x,y)$ were found to be negligibly small.
The fields $B_{\mathrm{rf}}(x,y,z)$ and $ B_{\mathrm{0}}(z)$ were each multiplied by a scaling factor, which detunes the static field or rf amplitude from the resonance condition, and the efficiency was simulated. The maximum efficiency averaged across the beam is shown in Fig. \ref{fig:ShoulderExample}.
This simulation imitates an experimental scan, in which the overall magnitude of each field can be tuned by the applied current but the relative inhomogeneities are fixed. As expected, the efficiency is much more sensitive to $ B_{\mathrm{0}}$ than to $ B_{\mathrm{rf}}$. As a check of the Bloch solver, a device with perfectly homogeneous $B_{\mathrm{0}}$ and $B_{\mathrm{rf}}$ was simulated and the efficiency was found to match the expected form given in Eqn. \eqref{eq:spin flip prob}.
The simulated maximum efficiency of a single flipper in flipping mode with 24 $\times$ 24 mm beam at 2 MHz was found to be 95.3\%. See Fig. \ref{fig:simbeamefficiency} for a 2D plot of efficiency vs pixel.

Our flipping-mode simulations also uncovered some general qualitative results. First, the exact profile of the rf envelope was found to not have a substantial effect on the maximum flipping efficiency; by using elementary time-dependent perturbation theory, one can see that the relevant parameter is only the integral of the rf envelope (see Appx. \ref{appx:perturbation}). The envelope does not have to be symmetric or have particularly sharp boundaries, but only be slowly varying compared to the rf oscillating field when transformed into the center-of-mass frame of the neutron.
The shape of the rf envelope does change the width of the tuning curve from Lorentzian to a more Gaussian shape without a decrease in maximum flipping efficiency; therefore, it may be beneficial to specifically design rf flippers with a gradual change in rf amplitude along the $x$-axis in order to use larger beam sizes.
Second, the tuning curves from these simulations are also useful to understand the experimental data discussed in the next section. For example, the static field inhomogeneity determines the asymmetry of the tuning curve: if the static field profile is concave down, then
the sub-peak shoulder of the tuning curve at the larger rf amplitude is diminished while the smaller-amplitude sub-peak is enhanced, and vice-versa if concave up.

Finally, we mention that preliminary simulations of our rf flipper design in precession mode, which suggest that the maximum efficiency is limited by the non-longitudinal components of the rf field. As discussed in the traditional spin echo literature \cite{Zeyen_1988,Zeyen_1996,Pasini_2015}, adding multiple short compensation coils superposed near the ends of the rf solenoid may enhance the precession-mode efficiency by reducing the strength of the radial rf fields.
In principle, the most important condition for a high-efficiency rf flipper in precession mode is that the polarization returns to the equator of the Bloch sphere, even if the desired azimuthal phase defined in Eqn. \eqref{eq:rf flipper phase} is not quite attained: the azimuthal phase can be adjusted by an additional weak static field while the introduction of a polar phase will typically lead to beam depolarization in the conventional NRSE setup.
Therefore, in analogy to a proposal for traditional NSE\cite{Piegsa_2016}, the most efficient parameter for an echo scan would be the phase of rf flipper $\phi_0$ in Eqn. \eqref{eq:lab Ham} as the polar phase change would be zero throughout the entire scan while the efficiency would be a simple linear function of rf phase.
Also, as the magnetic field envelopes are constant during the scan, potential issues due to hysteresis and coupling between the rf and static fields would be minimized.

\section{Flipping Mode Results}

\begin{figure}[t]
\includegraphics[width=.95\linewidth]{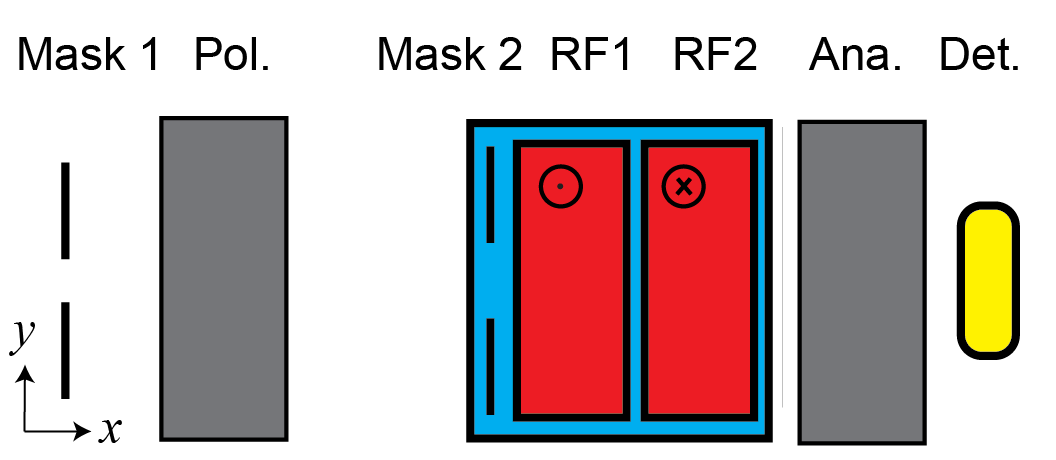}
\caption{\label{figBeamlineSetup} The experiment consisted of a polarizer, the two rf flippers (rf1 and rf2) in bootstrap mode, and an analyzer before the detector. The neutron beam travels from left to right. The beam collimation was determined by two masks, one before the polarizer and one attached to rf1. A weak vertical guide field ($\gtrsim~1$~mT) was maintained between the polarizer, device, and analyzer.}
\end{figure}

A test of the rf flipper performance at an effective frequency of 4 MHz was conducted at the beamline HB2D at the High Flux Isotope Reactor (HFIR) with a neutron wavelength of 0.43 nm. The two rf flippers (rf1 and rf2) in the single vacuum chamber were in bootstrap mode, meaning that the static fields were approximately the same strength but in opposite direction, so that they act as a single rf flipper with an effective frequency of the sum of the frequency of rf1 and rf2. \cite{KellerTbootstrap} Thus, in the 4 MHz experiment, rf1 and rf2 were each run at approximately 2 MHz. The rf flippers were placed between a neutron polarizer and analyzer, as shown in Fig. \ref{figBeamlineSetup}. The polarizer and analyzer were both $m = 4$ V-cavities produced by SwissNeutronics AG.
A 1~inch diameter He-3 tube detector with no position sensitivity was used. The detector background intensity was negligible. Several static guide fields were placed between the polarizer, rf flippers, and analyzer in order to maintain a strong field direction to keep a field of at least 1 mT along the entire beamline. The neutron entered the device polarization direction either parallel or anti-parallel with vertical static field in rf1.
Two masks were installed on the beamline, a $1.9 \times 1.1$~cm mask in the ``mask 1'' position shown in Fig. \ref{figBeamlineSetup} and a $2.5 \times 2.5$~cm mask at the ``mask 2'' position. These masks were separated by a distance of $\sim1$~m. 

The performance of the bootstrapped rf flipper was determined by measuring the neutron intensity of the beam as the static field was scanned through the resonance condition. Ideally, a perfectly-efficient flipper on a perfectly-polarized beamline should result in zero intensity at the resonance condition and the full beam intensity when the $B_0$ field is sufficiently far off resonance. If the initial polarization direction is reversed by reversing the guide field direction, then the resonance condition will have the full beam intensity and the intensity will be zero off resonance.
However, in practice the intensity of each curve is given by the combined efficiency of each individual rf flipper as well as the other beamline components, e.g. the polarizer, analyzer, and coupling of the guide field to the flipper's static fields. The efficiency of each individual rf flipper, as well as the polarization of the overall beamline, can be independently determined by a two flipper measurement which consists of the four permutations of the rf flippers being on or off. \cite{Felici1987,Wildes_1999} Here we assume the flipper efficiencies are constant.
The efficiency $f_i$ of each flipper is defined as the fraction of neutrons flipped when the flipper is tuned to the resonance condition.

Assuming that the rf flippers do not affect the polarization when turned off, the intensity with both rf flippers off is given by 
\begin{equation}
    N_{00} = I (1 + P_{B})
\end{equation}
where $I$ is the intensity scaling factor of the beamline that depends on the beam divergence, incident flux, and wavelength; and $P_{B}$ is the combined efficiencies of the polarizer, analyzer, and other components on the beamline, including the effects of depolarizing magnetic coupling between rf1 and rf2.
Defining the measured intensities with rf1 on, rf2 off as $N_{10}$, rf1 off, rf2 on as $N_{01}$, and both on as $N_{11}$, we find
\begin{subequations}
\begin{align}
    N_{01} =& I \big[ 1 + P_{B}(1 - 2f_1) \big] \\
    N_{10} =& I \big[ 1 + P_{B}(1 - 2f_2) \big] \\
    N_{11} =& I \big[ 1 + P_{B}(1 - 2f_2)(1 - 2f_1) \big]
\end{align}
\end{subequations}
where $f_1$ and $f_2$ are the efficiencies of rf1 and rf2, respectively.
Solving these equations for $f_1$, $f_2$, and $P_B$ gives
\begin{subequations} \label{eq:efficiencies}
\begin{align}
    f_1 &= \frac{1}{2} \left( 1 + \frac{N_{11} - N_{10} }{N_{00}- N_{01} }\right) \\
    f_2 &= \frac{1}{2} \left( 1 + \frac{N_{11} - N_{01} }{N_{00}- N_{10} }\right) \\
    P_{B} &= \frac{( N_{00} - N_{01})( N_{00}- N_{10})}{N_{00} N_{11} - N_{01} N_{10}}.
\end{align}
\end{subequations}

\begin{figure}[ht]
\includegraphics[width=.95\linewidth]{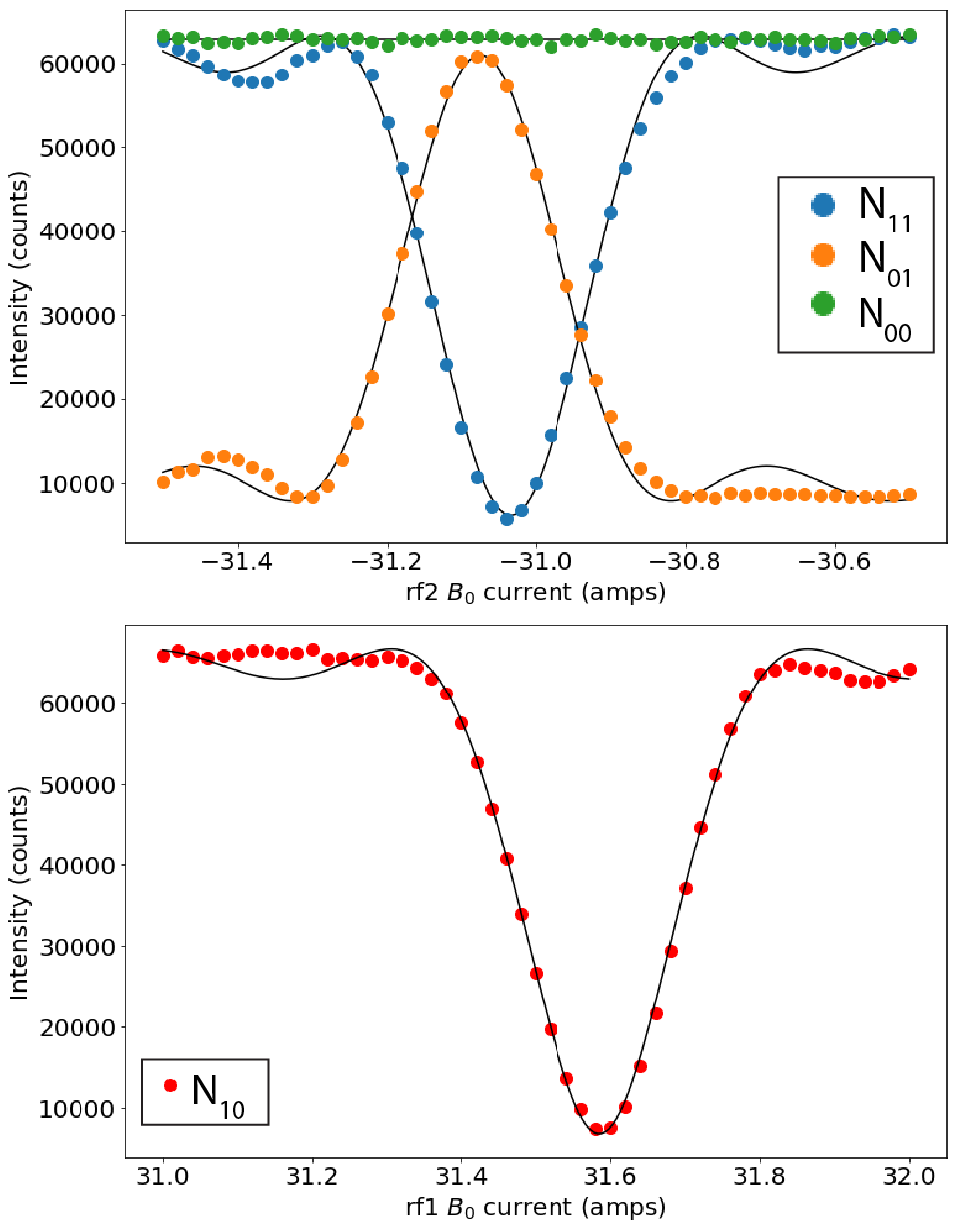}
\caption{\label{figFlippingModeData} (a) Observed neutron intensity as the current driving $B_0$ field of RF2 is scanned through the resonance condition with both rf coils off (N00), only rf2 on at 2 MHz (N01), and both rf1 and rf2 on at 2 MHz (N11). The black line is the expected curve for a perfectly homogeneous field, as given by Eqn. \eqref{eq:spin flip prob}. (b) The N10 intensity data with rf1 on at 2 MHz and rf2 off is shown as the current of the static-field of rf1 is driven through the resonance condition.}
\end{figure}

The values for $N_{00}, N_{01}, N_{10}$ and $N_{11}$ were found by scanning the $B_0$ field of rf1 or rf2 with the corresponding rf fields turned on. The results of these four curves can be seen in Fig. \ref{figFlippingModeData}. The $N_{10}$ intensity is shown in a separate plot because $B_0$ for rf1 was scanned, while for the other three curves, rf2 was scanned.
The flat curve for $N_{00}$ shows that it is a good assumption that a rf flipper will not cause a flip when turned off. Without fitting, we can see that the maximum intensity of the data in $N_{11}$ scan is $\sim60800$ and we can see that the average $N_{00}$ value $\sim62900$, so we expect a combined flipper polarization of approximately 97\%. 
To more accurately interpolate the peak intensities, we performed the standard Monte Carlo bootstrap procedure on the experimental data, using a non-parametric method via cubic splines.\cite{Press_1993} The result for the peak values from $10^6$ resamplings are shown in Tab. \ref{tab:bootstrap results}. The experimental result is an efficiency of 96.6$\pm 0.5\%$ for the bootstrap rf flipper at an effective frequency of 4 MHz.
The two flipper measurement at 4 MHz found a  polarization of $P_B = 83\%$ of the combined polarizer, analyzer and guide fields. The polarizer and analyzer each are expected to have a polarization efficiency of above 98$\%$, so it is likely that the coupling between the rf flippers and the rest of the beamline caused depolarization.
There are two regions where the stray field is most likely to cause depolarization: between the rf flippers and between each rf flipper and the vacuum chamber. The separation between the flippers is 5~mm and between the flipper outside face and the outside of the vacuum chamber is $\sim$2 cm. The stray field in these regions could be reduced by adding more mu-metal shielding.

\begin{table}[ht]
\centering
\newcolumntype{R}{>{\centering\arraybackslash}X}
\begin{tabularx}{.9\linewidth}{R|R|R|R}
    $N_{00}$ & $N_{10}$ & $N_{01}$ & $N_{11}$ \\
    \hline
    $60904 \pm 199$ & $7190 \pm 77$ & $5786 \pm 76$ & $62899 \pm 35$ \\
\end{tabularx}
\caption{\label{tab:bootstrap results} Experimental results of the intensities $N$ from the experimental data that are used to determine the rf flipper efficiencies $f_i$ and baseline beam polarization $P_B$ in Eqn. \eqref{eq:efficiencies}. The intensities are a result of an interpolation of the data in Fig. \ref{figFlippingModeData} using the Monte Carlo bootstrap method with $10^6$ resamplings.\cite{Press_1993}}
\end{table}

\begin{figure}[t]
\includegraphics[width=.95\linewidth]{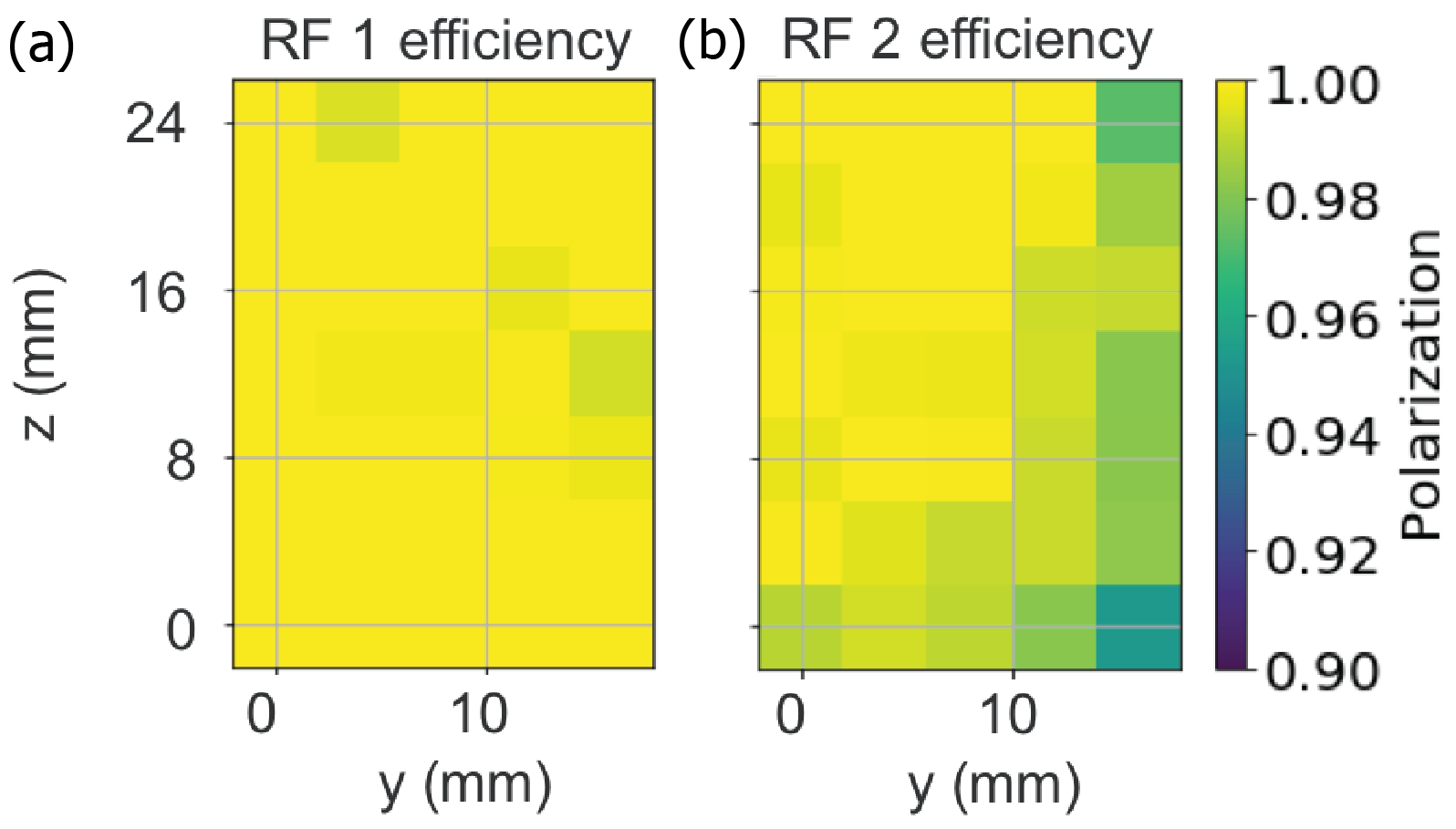}
\caption{\label{fig:2D efficiency} A 2D map of the flipping efficiency for each rf flipper, calculated from the data taken at 1.2 MHz and fit using Eqn. \eqref{eq:spin flip prob}. }
\end{figure}

We also measured the efficiency of flippers at a lower effective frequency of 2.4 MHz as using a 2D detector. This measurement was performed at the CG4B beamline at the HFIR with a neutron wavelength 0.55 nm and bandwidth of less than 0.005 nm. Both rf1 and rf2 were set to 1.2 MHz to give a bootstrap frequency of 2.4 MHz. The set-up was similar to the 4 MHz set-up except the analyzer was an s-bender and the detector used was an Anger camera with 2 dimensional pixels of size $\sim0.2$ mm which were binned into $4 \times 4$ mm pixels. \cite{LOYD_2024_Anger}. The beam shape is not rectangular because the Anger camera had a usable area of $24 \times 16$ mm in this set-up.
Masks of size $1 \times 1$ cm (mask 1) and $2.5 \times 2.5$ cm (mask 2) were separated by $\sim1$ m, such that the beam was mostly parallel to the optic axis.  Therefore, the pixel position on the detector largely corresponds to the position inside of the flipper. By performing the same two flipper measurement as above, the efficiency at each pixel was measured for both rf flippers.
Fig. \ref{fig:2D efficiency}, shows the efficiency for each pixel for both rf flippers. The high efficiency and uniformity suggests that the $B_0$ homogeneity is similar across the beam. However, there is an unexpected asymmetry in the rf2 data, with the lower right corner showing a few-percent lower efficiency. 

The flipping mode efficiency results and the Bloch solver simulations are in good agreement, within 1$\%$ for each rf flipper, demonstrating the homogeneity of the internal static field in the rf flippers and that the MagNet software simulates the internal field well. 

\section{MIEZE results} 

The rf flipper polarization efficiency was also measured in the so-called modulated intensity via zero effort (MIEZE) mode, a variation of NRSE in which only two rf flippers are needed. \cite{Felber_1999} This mode is a more sensitive probe of the internal field homogeneity of the rf flipper because the neutron is precessing through the rf flippers. Essentially, a MIEZE mode measurement consists of running rf1 and rf2 at different frequencies and measuring the time-oscillating neutron intensity at the detector. Theoretical details of this technique can be found in  Keller.\cite{Keller2002}
Each rf flipper will change the energy phase of the neutron proportional to its frequency, so the frequencies can be chosen based on a focusing condition to maximize the amplitude of the oscillating intensity signal at the detector position. The MIEZE frequency is the oscillation frequency of this signal and is twice the difference in the rf frequencies at the focal point.\cite{1998Besenbock,oda_tuning_2020} In this case, $\pi/2$ flippers are added after the polarizer and before the analyzer such that the neutron precesses throughout the device including through the uniform guide field.
This experiment was conducted at CG4B with a s-bender analyzer and a scintillator-based Timepix3 detector. \cite{Losko_2021,Funama_2024} For this measurement, rf1 was set to 50 kHz and rf2 to 100 kHz, and no matching circuits were used. The Fourier time of the measurement was 10 ps \cite{Franz2019}. An additional $1 \times 1$ cm mask was placed in front of the detector.

\begin{figure}[t]
\includegraphics[width=.95\linewidth]{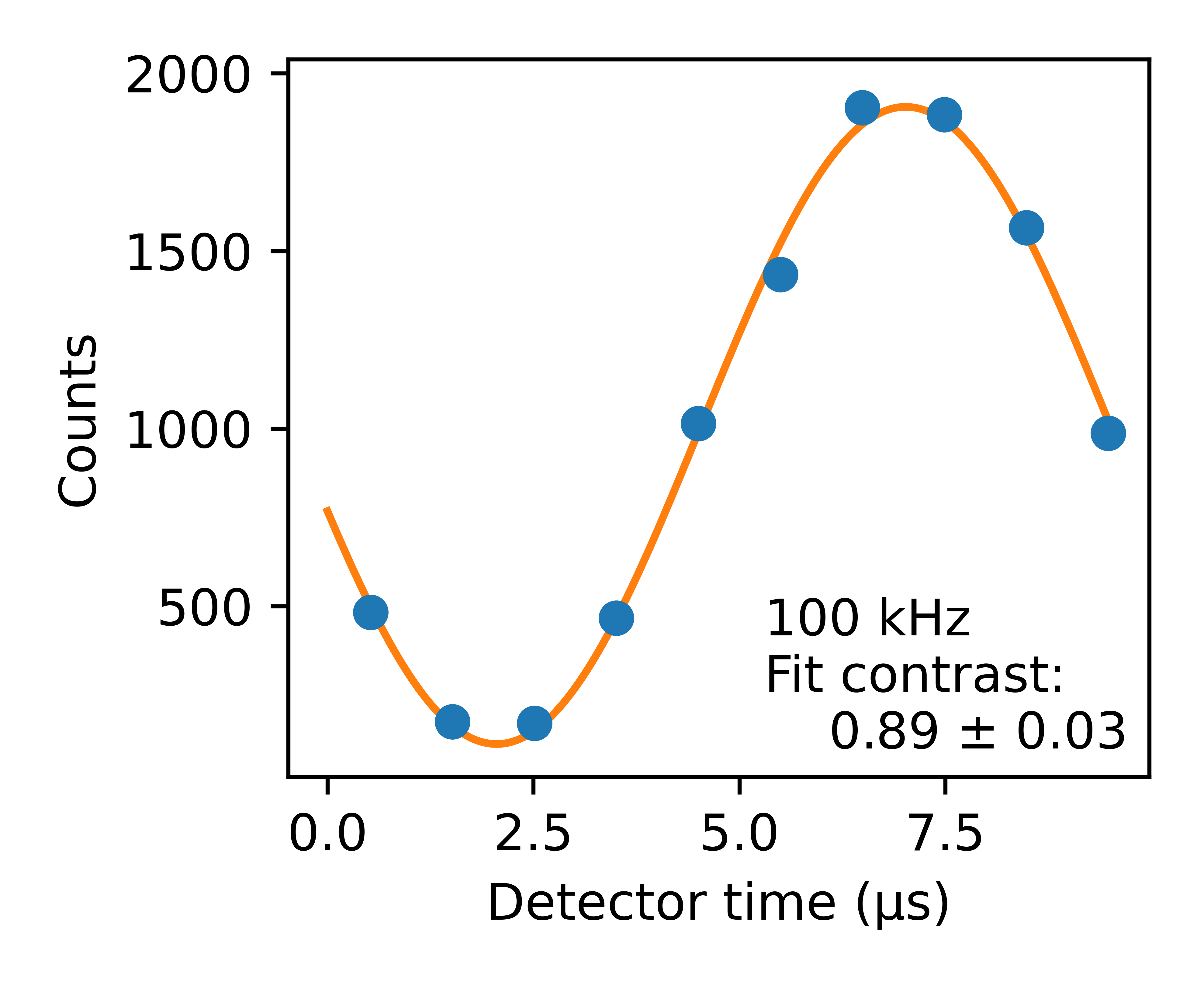}
\caption{\label{figMIEZE} The measured integrated intensity vs. detector time in MIEZE mode. Error bars are the size of the data points or smaller. The frequency is determined by the difference in the rf flipper frequencies. The detector time is reset at the expected MIEZE frequency.}
\end{figure}

The neutron intensity is shown in Fig.\ref{figMIEZE}, fitted to a cosine function. The fit contrast of 0.89$\pm0.03$ is comparable to the results of other high efficiency rf flippers at this frequency.\cite{Li2020} The fit frequency was 100.7 $\pm 1.3$ kHz, consistent with the expected MIEZE frequency of 100 kHz. 
However, the contrast is noticeably lower than the 96.6$\%$ efficiency seen in the flipping mode measurement. The efficiency of an rf flipper is typically higher than the MIEZE contrast for two primary reasons. The first is that, because both flippers need to be on to measure the MIEZE contrast, only the combined effect of the polarizers, guide field, rf flippers, analyzer, and detector can be measured.
The second reason is that in MIEZE mode, a difference in field integral will lead to a decreased contrast, while in flipping mode only the field direction matters. Thus, if the fields inside the rf flipper are pointed in an almost constant direction inside the device but have varying strength across the beam, then the flipping-mode efficiency will be high but the MIEZE contrast will be decreased. Therefore, it is difficult to determine if the contrast is lower than the efficiency due to the guide field homogeneity or due to the rf flipper field homogeneity. To maintain the MIEZE mode focusing, changing the frequency of the rf flippers also requires changing the guide field strength, making it complicated to decouple these effects.

\section{Discussion}

Our goal throughout the design process was to maximize the flipping efficiency of our rf flipper design without reducing the maximum spin echo time of $\tau_{\mathrm{NRSE}} \sim \SI{1000}{\nano \second}$ (also called the Fourier time). For an NRSE instrument, the spin echo time is given by
\begin{equation} \label{eq:spin echo time}
    \tau_{\mathrm{NRSE}} = \frac{2 m^2 \Delta L}{h^2} f \lambda^3,
\end{equation}
where $\Delta L$ is the distance between rf flippers.\cite{Keller2002} 
From Eqn.~\eqref{eq:spin echo time}, we see that to maximize the spin echo time, we must maximize rf frequency while keeping a reasonable flipping efficiency, which determines the contrast of the measured polarization.
In this section, we compare our design to other rf flippers and discuss future improvements to our design that maximize the spin echo time, which most practically means maximizing the rf frequency.

\subsection{Comparison with other rf flippers}
The device presented in this work is similar to the rf flipper designed by Li.\cite{Li2020} Both are bootstrap, transverse rf flippers with longitudinal rf fields, and both use HTS films as Meissner screens and HTS wire for high current. However, the Li rf flipper was designed for adiabatic mode and included a gradient field coil, making the efficiency in resonant mode lower (93.6$\%$ at 2 MHz). However, the efficiency in adiabatic mode is quite high (98$\%$ at 2 MHz), but the power output of $\sim$20 Watts and the need to cool the HTS films made the maximum frequency 2 MHz in bootstrap mode. The beam size and wavelength used in those experiments were approximately the same as in this flipper.

A key advantage of these types of rf flippers over static-field coils used in conventional NSE instruments and some other flipper designs is the lack of material in the beam. The only material in the beam is the gold-coated YBCO film on a sapphire substrate. The transmission through four of these devices for 1 nm neutrons would be $\sim 93\%$, with most of the absorption from the sapphire substrate.

The Reseda beamline at FRM-II uses two longitudinal, resonant RF flippers for MIEZE measurements.\cite{Franz2019} For these flippers, the static field is along the beam direction and the rf field is transverse. The maximum frequency tested has been 3.6 MHz. The beam size is $3.6 \times 3.6$ cm with a polarization of $\sim 92\%$.\cite{Jochum2020rf} Based on our experimental results, we expect our rf flippers would perform comparably in the same set-up, with a slightly higher frequency and a smaller beam size.
If four of our rf flippers were incorporated into an NRSE instrument, the field integral would be 0.55 Tm with a typical separation of the flippers of 2 meters.

A longtime concern in using transverse rf flippers has been the ability to correct for aberrations from different path lengths of neutrons through the instrument. Because much of the instrument is in zero-field, NRSE instruments typically need less correction from field aberrations than NSE instruments. However, neutrons travelling at different angles through the instrument will still acquire aberrant phases unless correction elements are added.
In the longitudinal rf flippers, Fresnel coils that are the same type as used for NSE can be used.\cite{Krautloher2016} Transverse-field correction methods have recently been developed using ellipsoidal mirrors \cite{FunamaF} or a prototype of a quadratic field integral phase correction coil.\cite{KuhnStephenJ2023RevSci} At Reseda, no correction elements are used for spin echo times up to 20 nanoseconds, suggesting that no correction elements would be required for Fourier times up to 20 nanoseconds regardless of the type of rf flipper.\cite{Franz2019}

\subsection{Future Improvements}
One way to increase the effective frequency of the device is to increase the number of bootstrap coils. A flipper with a fixed field inhomogeneity ratio $\Delta B_0$ will have a much lower efficiency when it is at higher frequency, as can be shown by a Taylor expansion of Eqn. \eqref{eq:spin flip prob} as discussed by Martin.\cite{Martin_2014}
This is also shown by the Bloch solver for the simulated field inhomogeneity of this device. With a 24 $\times$ 24 mm beam, the efficiency at 8~MHz is 69$\%$. Meanwhile, the probability of an effective 8 MHz flip from four 2 MHz flippers in bootstrap mode is $0.98^4 = 92\%$.  Bootstrapping the rf flippers leads to a significant increase in the polarization, even though the flippers have the same relative $\Delta B_0$. This is a significant advantage of using the transverse-field rf flippers that can be run in bootstrap mode. This result will be compounded by the two flippers required for a MIEZE beamline or four flippers require for an NRSE beamline.
Moreover, bootstrapping our rf flippers would enable higher effective frequencies in the future without pushing against the thermal limit.

The disadvantages of increasing the number of flippers are the increased material in the beam, device length, and calibration time. Our flippers only have thin HTS YBCO films deposited on a sapphire substrate in the beam, so material in the beam is the most negligible disadvantage. Adding more flippers in bootstrap mode will make the device longer, decreasing the divergence acceptance angle.
The largest difficulty is the tuning procedure. In this experiment, the optimum tuning parameters were found at each frequency by measuring the neutron flipping ratio. Practically, this would be too time-consuming as typical spin-echo experiments require many different frequencies to access different spin echo times. Adding more flippers in bootstrap mode exacerbates this problem. If a calibration pick-up coil for the rf field and Hall probe for the static field can be added close to the beam center without disrupting the rf resonance, the flipper may be able to be tuned using a pre-determined calibration value in those devices rather than using neutrons. In that case, the greatest obstacle for adding more bootstrap coils is the overall cost and reliability of each flipping device. 

If a no-neutron tuning technique is found, one can imagine the two bootstrap coils being replaced with four (or more) bootstrapped coils. In that case there would be 16 total rf flippers on an NRSE instrument and a large dynamic range can be accessed by turning off some of the rf coils and using them as static field subtraction coils, such as those between the RF flippers on the Reseda \cite{Jochum2020focus} or Larmor instruments.\cite{kuhngeeritswater2021} In principle, this enables an NRSE instrument using these flippers to reach very short spin echo times while maintaining high polarization.

The most direct improvement to a rf flipper for a resonant spin-echo instrument is to increase the maximum frequency of each rf coil. For our rf flipper, the maximum frequency with high efficiency was found to be 2 MHz for each coil. Above that frequency, the static field inhomogeneity decreases the efficiency.  This may be improved by shortening the length of the coil. By shortening the coil, the rf field magnitude needs to be increased in order to keep the same field integral. The higher rf field strength will reduce the depolarizing effect of static field inhomogeneities.\cite{Martin_2014}
This is more useful at higher frequencies where the static field inhomogeneity decreases the flipping efficiency more.
Even with a shorter coil, a higher frequency may be hard to achieve because eventually the static field will be strong enough to penetrate the superconducting YBCO films, reducing the sharpness of the field boundary.
Additionally, the rf field inhomogeneities will be more severe as the coil length is shortened, with the transverse rf components eventually limiting the flipper efficiency.

In order to maximize the flux on the sample, a large divergence acceptance angle is preferred for a RF flipper. This can be achieved by increasing the beam size through the device. Simulation results show that the rf field homogeneity will decrease noticeably as the diameter is increased, so this would require a more sophisticated coil design to correct for these aberrations.
A way to increase the divergence acceptance angle without increasing the beam size is to add neutron guides between the rf flippers, as the Reseda beamline already does. The correction of the phase due to path length aberrations between the rf flippers is still possible with neutron guides, although not for the entire length.\cite{kuhn2021}

Because the spin echo time is proportional to $\lambda^3$ [see Eq. \eqref{eq:spin echo time}], a slight increase in the wavelength will greatly increase the spin echo time. However, a longer wavelength hurts the instrument performance in two ways. First, the longer wavelength will reduce the flux. At the cold neutron source at the HFIR, the flux is proportional to $\lambda^{-4}$ for wavelengths above 0.6 nm.\cite{Robertson2008} Second, to keep the rf flipper in resonance, as the wavelength is increased, the rf field strength is decreased. This will make the device more sensitive to static field inhomogeneities, as discussed above.

Even without any of the improvements discussed above, the existing rf flipper is well-suited to a MIEZE-type instrument. As discussed above, the rf flipper is comparable in frequency to those currently installed at the Reseda and Larmor beamlines. There are straightforward options to improving our existing instrument, largely because the transverse-field enables four bootstrapped coils as a way to double the frequency. The Bloch solver developed for this paper matched the experimental results well, so a new version of the rf flipper can be simulated and optimized before being constructed.

Therefore, a 4 bootstrap, effective 8~MHz flipper can be realized, then a 1000~ns spin echo time NRSE instrument could be built. More specifically, an instrument with the flippers set 4 m apart would provide a field integral of 2.2~Tm, which gives a spin echo time of 100~ns with 0.63~nm neutrons and 1000~ns with 1.35~nm neutrons. With the polarization at 96.6$\%$ for each 4~MHz flipper, equivalent to what was found in flipping mode at 4~MHz in this measurement, the overall efficiency of the flippers would be $\sim 75~\%$. 
This assumes that the efficiency can be maintained at wider bandwidths and up to 1.35~nm as well as that the contrast is close to the overall efficiency. These assumptions have proven to be true in other spin echo instruments. If the contrast can be kept close to the overall efficiency, this would make a competitive instrument for high-resolution neutron spectroscopy. 

\section{Conclusion}

We have constructed a transverse, resonant rf flipper suitable for an NRSE type instrument. The resonance mode operation allows for low power output, which enables the flipper to be run at 4 MHz with 96.6$\pm 0.6\%$ efficiency in bootstrap mode. To our knowledge, this is the highest effective operating frequency tested for a neutron spin flipper.
We have also presented a Bloch solver to simulate the efficiency of rf flippers with non-uniform static or radio-frequency fields and will use this solver along with finite-element magnetic field simulations to further reduce the field inhomogeneity of future designs. With this method of understanding the inhomogeneity, we maintain that we can design a resonant rf flipper suitable for a high-performance NRSE instrument.

\FloatBarrier
\section{Acknowledgements}
The authors would like to thank Lowell Crow and Rob Dalgliesh for useful discussions. Machining was primarily done by the Indiana University Physics machine shop: John Frye, Danny Clark, Darren Nevitt, and Todd Sampson. We thank Matthew Loyd for assistance with the Anger camera.

This research used resources at the High Flux Isotope Reactor, a DOE Office of Science User Facility operated by the Oak Ridge National Laboratory. F. Li would also like to acknowledge the support from DOE Early Career Research Program Award (KC0402010), under Contract No. DE-AC05- 00OR22725.

The work reported here was funded by the Department of Energy STTR program under grants DE-SC0021482 and DE-SC0018453.
S. McKay acknowledges support from the US Department of Commerce through co-operative agreement number 70NANB15H259.

\appendix
\section{Quantum Bloch solver theoretical methods} \label{appx:qbs}

In this appendix, we derive the equations used in the quantum Bloch solver used to simulate the rf flipper, including a discussion of the necessary assumptions.
Starting with the time-dependent Schrodinger equation $i \hbar \partial_t \ket{\psi(t,\bm{r})} = \mathcal{H}(t,\bm{r}) \ket{\psi(t,\bm{r})}$, where $\ket{\psi(t,\bm{r})}$ is the neutron spinor and $\mathcal{H}(t,\bm{r})$ the single-neutron Hamiltonian, we can write down the formal solution in terms of the time-evolution operator: $\ket{\psi(t,\bm{r})} = \mathcal{U}(t,t_0) \ket{\psi(t_0,\bm{r})}$ where
$\mathcal{U}(t,t_0)=\mathcal{T} \exp\left[ -\frac{i}{\hbar} \int_{t_0}^t dt' \,\mathcal{H}(t', \bm{r}) \right]$ with $\mathcal{T}$ as the time-ordering operator. The time-ordering operator is needed as the direction of the magnetic field changes, and thus the Hamiltonian at different times does not commute with itself.

The first approximation we take is to assume that the kinetic and potential terms of the time-evolution operator can be expressed in the following way:
\begin{equation}
\begin{aligned}
    \mathcal{U}(t,t_0) \approx&  \exp \left( - \frac{i}{\hbar} \int_{t_0}^t dt' \, \frac{\bm{p}^2}{2m} \right) \times \\
    &\mathcal{T} \exp \left( \frac{i}{\hbar} \int_{t_0}^t dt' \, \bm \mu \cdot \bm B(t',\bm r) \right).
\end{aligned}
\end{equation}
By using this approximation, we are assuming that the neutron travels the ``classical'' path determined by Ehrenfest's theorem; we also assume pure transmission of the neutron through the region of non-zero magnetic field as the reflection coefficient is vanishingly small for the considered field magnitudes. Although the path that the neutron takes is treated classically in this approximation, the spin state is still treated quantum mechanically as a two-level system (or qubit).
We call this particular factorization the \textit{semiclassical approximation} as it corresponds to a semiclassical view of the neutron where the neutron polarization precesses with a frequency $\gamma \bm{B}$ about the direction of the applied magnetic field.
Higher-order corrections depend on the commutator of the kinetic and magnetic terms given by the Zassenhaus formula, a relative to the well-known Baker-Campbell-Hausdorff and Suzuki-Trotter formulas.\cite{Suzuki_1976}

As a second assumption, we will only consider planewave solutions.
Each neutron spin state is more accurately described by a finite-size wavepacket which can spatially separate, an example of the single-particle Stern-Gerlach effect.\cite{Gerlach_Stern_1922,Sherwood_1954} However, as long as the longitudinal and transverse coherence lengths of the neutron are larger than the spatial separation, each component of the spinor can typically be treated as a planewave; if the separation is larger than the coherence lengths at the detection time, one would observe anomalous depolarization as the two wavepackets would no longer be significantly overlapping.\cite{Arend_2004}
For the magnetic fields we consider in this paper, this spatial separation is at most 1 micron at a rf frequency of \SI{4}{\mega\hertz}. The coherence length of the neutron is has been observed to be of order of microns or larger, so we are safely in the planewave limit.\cite{Wagh_2011,Majkrzak_2022,McKay_2024}
From the well-known duality between planewaves and rays,\cite{Goldstein_2011} we can now treat the kinetic portion of the time-evolution operator as a classical ray, and so we restrict ourselves to calculating the spin portion of the time-evolution in the center-of-mass frame of the neutron.
We also ignore the Doppler effect when we use the usual Galilean transform $(x,t)_{\mathrm{lab}} \to (x-vt,t)_{\mathrm{com}}$ to shift from the lab frame into the neutron's center-of-mass frame as the frequency correction is only around 0.1\%  even for an rf frequency of 4 MHz.
Therefore, with these approximations, we have transformed the problem into an ordinary differential equation in time only.

Our goal is now to numerically solve the equation
\begin{equation}
    i \hbar \frac{d}{dt} \ket{\psi(t)} = \mathcal{H}(t) \ket{\psi(t)},
\end{equation}
with the choice $t_0 = 0$. Using the Magnus expansion \cite{Magnus_1954,Blanes_2009} for the $\bm B$-dependent part of the time-evolution operator $\mathcal{U}^{(B)}$ in the semiclassical approximation, for a small enough time interval $\tau$ we find
\begin{equation}
    \mathcal{U}^{(B)}(t+\tau,t) \approx \exp \left( \frac{i}{\hbar} \int_{t}^{t+\tau} dt' \, \bm \mu \cdot \bm B(t',\bm r(t')) \right),
\end{equation}
where now the magnetic fields are strictly only a function of time as we have chosen to work in the center-of-mass frame of the neutron.
The Magnus expansion formalizes the intuitive idea that we can choose the time step small enough such that the Hamiltonian appears to be time-independent, which corresponds to the \textit{average Hamiltonian theory} in the nuclear-magnetic resonance literature.\cite{Mananga_2011} Although more accurate higher-order commutator expansions exist (e.g., the Magnus-Floquet, Fer, and Wilcox expansions) the first-order average Hamiltonian theory was found to be sufficient for our analysis.

The analytical form for our propagator can found by applying Euler's formula for quaternions, resulting with
\begin{equation} \label{eq:su2_prop}
\begin{aligned}
    \mathcal{U}^{(B)}(t+\tau,t) =& \mathbb{I}_2 \cos(\gamma |B| \tau / 2) - \\
    & i \, \hat{n} \cdot \sigma \sin(\gamma |B| \tau /2),
\end{aligned}
\end{equation}
where we have defined $\mathbb{I}_2$ as the $2 \times 2$ identity matrix and the magnetic field as $\bm{B}(t) = |B| \hat{n}$. More advanced magnetic field interpolation schemes were considered such as the method described by de Haan\cite{Haan_2013}, but we found the simple linear interpolation to be sufficient.
Finally, the neutron spinor at time $t = \tau (n+1)$ for time step $n \in \mathbb{N}$ is then given by the simple recursive equation $\ket{\psi_{n+1}} = \mathcal{U}^{(B)}_n \ket{\psi_n}$.

A convenient closed form for the propagator also exists for the polarization, known as Rodrigues' rotation formula:
\begin{equation} \label{eq:so3_prop}
\begin{aligned}
    \mathcal{U}_P^{(B)}(t+\tau,t) =& \mathbb{I}_3 + \hat{n} \cdot \Omega \sin(\gamma |B| \tau / 2) \\
    &+ (\hat{n} \cdot \Omega)^2 [1 - \cos(\gamma |B| \tau /2)].
\end{aligned}
\end{equation}
The matrix components of the vector $\Omega =(\Omega_x,\Omega_y,\Omega_z)$ are given by
\begin{equation}
\begin{aligned}
    \Omega_x = &\begin{bmatrix}
    0 & 0 & 0 \\
    0 & 0 & -1 \\
    0 & 1 & 0 
\end{bmatrix}, \quad
\Omega_y = \begin{bmatrix}
    0 & 0 & 1 \\
    0 & 0 & 0 \\
    -1 & 0 & 0 
\end{bmatrix}, \quad \\
&\qquad \Omega_z = \begin{bmatrix}
    0 & -1 & 0 \\
    1 & 0 & 0 \\
    0 & 0 & 0 
\end{bmatrix},
\end{aligned}
\end{equation}
which are the familiar generators of 3D rotations and $\mathbb{I}_3$ the $3 \times 3$ identity matrix. See Curtright\cite{Curtright_2015} for a closed form for the propagator for any spin value.
Notice that the spinor (polarization) propagator is unitary (orthogonal), so the probability (norm) of the state is conserved at every step, which enhances the numerical stability. Such numerical solutions are called \textit{geometric integrators} as they preserve the geometric structure that determine the flows of the governing differential equation.\cite{Olver_1993,Hairer_2006}
See the following references for discussions of the numerous computational advantages of the approach outlined here.\cite{Mananga_2016,Leskes_2010,Takegoshi_2015,Kais_2014}
For further insights and discussion of the connection between 3D rotations and the dynamics of qubits, see the works by Berry\cite{Berry_Robbins_1993} and Rojo.\cite{Rojo_2010}
We note that Mathews \textit{et al.} have also applied a similar numerical method in their spin tracking software which was developed for the neutron electric dipole moment experiment at the Spallation Neutron Source (nEDM@SNS).\cite{Mathews_2024}

Finally, we note that \textit{any} Bloch solver method does not exactly describe the time-evolution of the neutron as it does not account for the energy exchange between the rf field and neutron. To incorporate this change in energy, one can either treat the rf field classically or quantum-mechanically, as discussed in more detail in a series of works by Sulyok.\cite{Sulyok_2012,Sulyok_Lemmel_2012,Sulyok_2015}
A recent paper by Sobolev\cite{Sobolev_2023} showed mathematically that it is always possible to achieve a 100\% efficient $2 \pi$-flip with any arbitrary rf envelope, while the same was not found to be true for a $\pi$-flip, although they did not construct a rigorous proof of this efficieny limitation of the $\pi$-flip case.
A $2 \pi$-flip would result in an exchange of $2 \hbar \omega$ between the rf field and the neutron, increasing the spin echo time by another factor of 2 at the potential cost of increased field inhomogeneity.
Therefore, it may be useful to model and experimentally verify the efficiency of multi-photon exchange\cite{Muskat_1987,Summhammer_1995} with our rf flipper design.

\section{General resonance condition} \label{appx:perturbation}

Here we consider the case where the rf envelope is non-uniform using first-order time-dependent perturbation theory. We now show that the neutron spin flip probability for a non-uniform rf envelope simply depends on integral of the envelope.
For simplicity, we only consider magnetic fields of the following form:
\begin{equation}
    \bm{B}(t,\bm{r}(t)) = B_{\mathrm{rf}} \, g(t) \cos(\omega t)\hat{x} + B_0 \hat{z}
\end{equation}
where $B_{\mathrm{rf}}$ and $B_0$ are constants and $g(t)$ is an arbitrary envelope function that approaches zero at the flipper boundaries $t=0$ and $t=T$.
If one considers the rf field as a perturbation to the much stronger static field, then without loss of generality, the probability amplitude for a spin flip $\ket{\uparrow} \to \ket{\downarrow}$ is given by
\begin{equation}
    c_{\uparrow,\downarrow} \approx \frac{-i \gamma B_{\mathrm{rf}}}{2} \int_0^T dt' \, g(t') \cos(\omega t') e^{i \omega_b t'}
\end{equation}
where $\omega_b = (E_{\uparrow} - E_{\downarrow})/\hbar = \gamma B_0$ the usual Born frequency due to the uniform static field $B_0$. In the rotating wave approximation, the spin flip probability is given by
\begin{equation}
    P_{\uparrow,\downarrow} = |c_{\uparrow,\downarrow}|^2 \approx \left( \frac{\gamma B_{\mathrm{rf}}}{4} \right)^2
    \left| \int_0^T dt' \, g(t') e^{i(\omega_b - \omega)t'} \right|^2.
\end{equation}
We can immediately see that the location of the peak of the envelope only contributes an irrelevant phase factor.
For the idealized case where $g(t') = 1$ if $t' \in [0,T]$ and $g(t') = 0$ otherwise, we find
\begin{equation}
    P^{\mathrm{(id)}}_{\uparrow,\downarrow} = \left( \frac{\gamma B_{\mathrm{rf}} T}{4} \right)^2 \sinc^2\left( \frac{\omega_b - \omega}{2} T \right)
\end{equation}
which agrees with Eqn. \eqref{eq:spin flip prob} when sufficiently off-resonance.

Considering the limit of the ratio of the ideal result to the general case as the detuning parameter $\omega_b-\omega$ goes to zero, we obtain the promised result, namely the squared ratio of the area of the envelopes: 
\begin{equation}
    \lim_{\omega \to \omega_b} \frac{P^{\mathrm{(id)}}_{\uparrow,\downarrow}}{P_{\uparrow,\downarrow}} = \frac{T^2}{|\int dt' \, g(t')|^2}.
\end{equation}
Therefore, to ensure a spin flip, one has to multiply the rf amplitude $B_{\mathrm{rf}}$ for the general envelope $g(t)$ by the square root of this ratio.
Although it is well-known that the first-order theory breaks down when approaching resonance,\cite{Gottfried_2003} if one considers the higher order terms in the Dyson series and takes the resonance limit, each higher-order coefficient will contain a a similar integral over $g(t)$, meaning that this first-order behavior still holds at resonance as confirmed numerically by our Bloch solver for a variety of functions, including non-symmetric and multimodal functions.

\bibliographystyle{apsrev4-2.bst}
\bibliography{RFbib.bib}
 
\end{document}